\documentclass[12pt]{article} 

\usepackage[utf8]{inputenc} 

\usepackage{color} 

\usepackage{graphicx} 
\usepackage{hyperref}
\hypersetup{
    colorlinks=true,
    linkcolor=blue,
    citecolor=red,
    filecolor=magenta, 
    urlcolor=cyan
}

\usepackage{mathrsfs}
\usepackage{psfrag,epsf}
\usepackage{enumerate}
\usepackage{url} 
\usepackage{booktabs} 
\usepackage{array} 
\usepackage{paralist} 
\usepackage{verbatim} 
\usepackage{subfig} 

\usepackage{fancyhdr} 
\usepackage{natbib}

\usepackage{amsmath,amsthm,amssymb} 
\usepackage{bbm}
\usepackage{tikz}
\usetikzlibrary[positioning,calc,decorations.pathreplacing]

\usepackage{xr}
\makeatletter
\newcommand*{\addFileDependency}[1]{
  \typeout{(#1)}
  \@addtofilelist{#1}
  \IfFileExists{#1}{}{\typeout{No file #1.}}
}
\makeatother

\graphicspath{{Plot/}}

\newtheorem{lemma}{Lemma}
\newtheorem{theorem}{Theorem}

\newtheorem{ass}{Assumption}

\newtheorem{corollary}{Corollary}

\newtheorem{definition}{Definition}

\usepackage[doublespacing]{setspace}

\usepackage[nottoc,notlof,notlot]{tocbibind} 
\usepackage[titles,subfigure]{tocloft} 

\def\blue#1{\textcolor{blue}{#1}}
\def\red#1{\textcolor{red}{#1}}

\DeclareMathOperator*{\argmax}{arg\,max}

\def\mA{\mathcal{X}}
\def\mB{\mathcal{B}}
\def\mF{\mathcal{F}}
\def\mG{\mathcal{G}}
\def\bG{\boldsymbol{G}}
\def\mL{\mathcal{L}}
\def\bP{\boldsymbol{P}}

\def\xst{\tau}
\def\Xst{\mathtt{T}}
\def\XXst{\mathcal{T}}
\def\Re{\mathbb{R}}

\newcommand{\blind}{0}

\begin{document}

\def\spacingset#1{\renewcommand{\baselinestretch}%
{#1}\small\normalsize} \spacingset{1}


\if0\blind
{
  \title{\bf Order Statistics Approaches to Unobserved Heterogeneity in Auctions}
  \author{Yao Luo\thanks{
    \textit{Contact Information}: Luo: Department of Economics, University of Toronto, Max Gluskin House, 150 St.\ George St, Toronto, ON M5S 3G7, Canada (email: yao.luo@utoronto.ca); Sang: Department of Statistics and Actuarial Science, University of Waterloo, 200 University Avenue West, Waterloo, ON, Canada, N2L 3G1, Canada (email: psang@uwaterloo.ca); Xiao: Department of Economics, Indiana University, 100 S. Woodlawn Ave. Bloomington, IN 47405 (email: rulixiao@iu.edu). We thank the editor, an associate editor, anonymous referees, Yingyao Hu, and Ji-Liang Shiu for their time and helpful comments, and Qingyang Zhang for valuable research assistance. Luo acknowledges funding from the SSHRC Insight Grant. }\hspace{.2cm}\\
    Department of Economics, University of Toronto\\
    and \\
    Peijun Sang \\
    Department of Statistics and Actuarial Science, University of Waterloo\\
    and \\
    Ruli Xiao \\
    Department of Economics, Indiana University}
  \maketitle
} \fi

\if1\blind
{
  \bigskip
  \bigskip
  \bigskip
  \begin{center}
    {\LARGE\bf Order Statistics Approaches to Unobserved Heterogeneity in Auctions}
\end{center}
  \medskip
} \fi

\bigskip
\begin{abstract}
We establish nonparametric identification of auction models with continuous and nonseparable unobserved heterogeneity using three consecutive order statistics of bids. We then propose sieve maximum likelihood estimators for the joint distribution of unobserved heterogeneity and the private value, as well as their conditional and marginal distributions. Lastly, we apply our methodology to a novel dataset from judicial auctions in China. Our estimates suggest substantial gains from accounting for unobserved heterogeneity when setting reserve prices. We propose a simple scheme that achieves nearly optimal revenue by using the appraisal value as the reserve price.
\end{abstract}

\noindent%
\vfill

\newpage
\spacingset{1.8} 

\section{Introduction}

In empirical auction analysis, to estimate bidder value distributions, the analyst usually needs to pool data from auctions for similar but not identical items. However, the data available for these auctions often lack a precise description of the auctioned item. This situation results in auction-level unobserved heterogeneity (UH), which leads to inaccurate estimates of bidder value distributions and, therefore, misleading policy implications. For instance, \cite{HQT2019} finds that UH accounts for two-thirds of price variation after controlling for information provided in the eBay Motors auctions, and that ignoring this feature would dramatically mis-estimate the welfare measures. The existing literature adapts measurement error approaches to tackle such an issue. Suppose the analyst observes all bids. The analyst could then identify the value distribution using observed bids as measurements for the unobserved characteristics, since these bids are independent conditional on such unobserved characteristics.

However, the conditional independence condition fails when the analyst only observes incomplete bid data. This could occur for various reasons. First, in English or ascending outcry auctions, the bidder with the highest value only needs to outbid the bidder with the second-highest value to win, which means the recorded bids do not contain the highest value. Moreover, even in first-price sealed-bid auctions, where all bids are supposed to be submitted to the auctioneer, the auctioneer may still not record all the bids in practice: sometimes the auctioneer only records the most competitive bids, such as the top three bids in regular auctions or apparent low bids in procurement auctions. Thus, the econometrician can only observe a few order statistics of the bids, i.e., incomplete bid information. For instance, the U.S. Forest Service timber auctions only record at most the top 12 bids regardless of the number of bidders. The Washington State Department of Transportation provides an online archive of bid opening results that are six months or older, but only for the top three apparent low bids. Even if the auctioneer records all bids, the most competitive bids are often more accessible to the public. For instance, The Federal Deposit Insurance Corporation resolves insolvent banks using first-price auctions but only publishes the top two bids and bidders' identities \citep{allen2019resolving}. The three apparent low bids are one-click downloadable on the website of the California Department of Transportation. These order statistics are naturally dependent, invalidating conventional identification strategies. 



We make three contributions in this paper. First, our paper is the first to study identification of auction models with \textit{continuous and nonseparable} UH using incomplete bid data. Our specification allows for flexibility in how UH affects both bidder value and the equilibrium bidding strategy, i.e., the mapping from a bidder's private value to his/her bid.\footnote{Even if one assumes separable UH in the value, separability passing to the bid often requires additional institutional features or assumptions. See, e.g., \citet{andreyanov2020secret}.} 

Our identification strategy adapts \cite{hu2008instrumental} for nonclassical measurement error models to the auction setting. This extension is nontrivial in that we only observe order statistics of UH-contaminated bids. As a result, we cannot achieve a parsimonious conditional independence structure as in their work.\footnote{They assume that the outcome variable is independent of the observed independent variable and an instrument conditional on the unobserved true regressor.} Instead, we follow \cite{luo2020identification} and consider the most common case of incomplete bid data: \textit{consecutive} order statistics of bids. Their main insight is that consecutive order statistics have a semi-multiplicatively separable joint distribution with a simple indicator function capturing the correlation. Unlike both papers using two measurements with an instrument, we use \textit{three} consecutive order statistics of bids. Given a partition on the range of the measurements, we again obtain a separable structure traditionally achieved under conditional independence. This turns the identification problem into an operator diagonalization problem, allowing constructive identification arguments using linear operator tools. Moreover, we use these tools differently by considering bounded linear operators defined on a Hilbert space and taking values in another Hilbert space. This space is smaller than the $\mathcal{L}^1$ space adopted in \cite{hu2008instrumental}, which focuses on a Banach space. While we could also work with Banach space, using Hilbert space simplifies the analysis of relevant operators and thus our proofs thanks to many existing theoretical results.\footnote{For instance, it is straightforward to define the adjoint operator by using the concept of inner product in Hilbert spaces.}

Second, we propose sieve maximum likelihood estimators (MLE) of the model primitives and provide conditions that guarantee their consistency. The estimation of auction models allows for counterfactual policy analysis, such as computing the optimal reserve price. If UH is common knowledge among agents in the auction, it is a critical control in policy analysis. Therefore, optimal policy recommendation requires estimating the joint distribution of UH and bidder private value.\footnote{Since there is a known mapping between the bid distribution and the value distribution, we will use the two terms interchangeably. See \citet{GPV2000} and \citet{AH2002} for this mapping.} In particular, we approximate the joint density of bids and UH using the tensor product of two univariate sieve bases. We then represent the marginal density of the UH and the conditional distribution of the value using the sieve-approximated joint distribution. Therefore, these distributions are all estimated nonparametrically.\footnote{In contrast, previous research only focuses on the estimation of the joint distribution using a semiparametric structure \citep{chen2006efficient} and \citep{hu2008instrumental} or a nonparametric structure \citep{wu2012partially}.} \cite{hu2008instrumental} proposes sieve approximations to the conditional distribution and marginal distribution. Our sieve approximation to the joint distribution is more convenient as we just need to impose the normalization assumption on the joint distribution approximation once. 

The consistency of our estimator relies on the condition that the sieve space approximates well the joint distribution of bids and UH. To formalize this intuition, we quantify the complexity of this space using bracket entropy and prove consistency of the sieve MLEs for the joint, conditional, and marginal densities. We establish a concentration inequality based on the bracketing number, a similar notation to covering numbers used in \cite{hu2008instrumental}.  The online supplement Section S.2.2 further investigates the properties of B-splines and Bernstein polynomials, both of which are popular in empirical applications. 



Lastly, we apply our identification and estimation method to a novel dataset from judicial auctions conducted by a municipal court in China. By default, this court uses 70\% of the appraisal value as the starting price, which also serves as a reserve price.  Our estimation results suggest substantial gains from accounting for UH when designing reserve prices. The court can gain $5.81\%$ more revenue using an optimal reserve price for each item. However, this scheme is complex; the seller would need to know UH and recover the conditional density of bidder values. Instead, we propose a simple scheme that achieves nearly optimal revenue by using the appraisal value as the reserve price. Specifically, using the estimated model, we find that using the appraisal value as the reserve price achieves $98.85\%$ of the potential gains from the optimal reserve prices.

\subsection*{Literature Review}

The auction literature has widely applied techniques developed in the measurement error literature for identifying auction models with UH. If the UH is continuous and has a separable structure on bidder valuations, identification relies on the deconvolution approach and requires two random bids for each auction. See \citet{LV1998}, \citet{li2000conditionally}, and \citet{K2011}, among others. If the UH is finite and discrete, which by nature is nonseparable, identification relies on the condition that the bids are independent conditional on the UH and requires three random bids for each auction. See \citet{H2008}, \citet{HMS2013}, and \citet{Luo2018}.

Moreover, the literature has seen rapid growth in identifying and estimating auctions models using order statistics of bids. \citet{AH2002} shows that symmetric independent private value (IPV) auctions are identifiable by the transaction price and the number of bidders using the one-to-one mapping between the distribution of an order statistic and its parent distribution; \citet{komarova2013new} identifies asymmetric second-price auctions using the winner's identity and the transaction price; \citet{GL2018} shows that IPV first-price auctions without observable competition is identifiable using the transaction price; \cite{menzel} studies large sample properties for nonparametric estimators using order statistics of bids.

A growing literature tackles the identification of auction models with UH and incomplete bid information. Assuming the UH is finite and discrete, \citet{M2017} provides identification results from (any) five order statistics to restore the conditional independence condition by the Markov property of order statistics. \citet{luo2020identification} provides an alternative identification strategy using two \textit{consecutive} order statistics of bids and an instrument. Finiteness simplifies their identification arguments because model restrictions can be written in matrix algebra. In contrast, we use linear operators, which is not a trivial extension of the matrix operations. Moreover, we extend our identification results to allow for binding reserve prices and apply them in our empirical application.

In the framework of additively separable continuous UH, \citet{HQT2019} achieves point identification using English auction models, assuming piecewise real analytic density functions and using variations in the number of bidders across auctions; \citet{FL2017} provides identification results for ascending auctions, relying on reserve prices and two order statistics of bids. \citet{CLX2022} studies deconvolution using two order statistics. Our paper is the first to show point identification of auction models with continuous and nonseperable UH using incomplete bid data.


The remainder of this paper is organized as follows. Section \ref{sec:id} presents our main identification results. Section \ref{sec:sieve} proposes sieve maximum likelihood estimators. Section \ref{sec:app} presents an application to judicial auctions in China. Section \ref{sec:conclusion} concludes. The online supplement contains detailed proofs for the identification results and the asymptotic properties and the finite-sample properties of the proposed estimators.

\section{Main Identification Results} \label{sec:id}

Consider a standard IPV auction model for items with a scalar heterogeneous characteristic $\Xst$ that is observable to all bidders but unobserved to the analyst.
For simplicity, we abstract  from observable (to the analyst) characteristics. Suppose $n \geq 2$ symmetric bidders participate in an auction with zero reserve price.\footnote{We assume the number of potential bidders is known. Otherwise, we can treat it as an additional dimension of UH, as in \cite{luo2020identification}, or construct it through alternative data sources. In procurement auctions, we can construct it using the number of qualified firms in the local market via public information, such as the list of qualified firms and their contact information.} All bidders observe the characteristics $\Xst$ before they submit bids.\footnote{While we focus on regular auctions here, our results extend trivially to procurement auctions.} Our identification strategy applies regardless of whether the seller observes $\Xst$ or not. Among $n$ potential bidders, bidder $i$, where $i=1,...,n$, draws his/her value $V_i$ from the conditional value distribution $f^{V|\Xst}(v|\xst)$ and submits a bid $X_i$. We consider the situation wherein the latent auction characteristic and bids/values are continuous.  We denote the marginal distribution of the latent characteristic $\Xst$ as $f^{\Xst}(\xst)$ and the optimal conditional bid distribution as $f^{X|\Xst}(x|\xst)$, where $x$ is the optimal bid. 

We first introduce the standard assumption regarding the value distribution.
\begin{ass} \label{cond_id} (Conditional Independence) Bidder values, $V_1$,..., $V_n$, are i.i.d. conditional  on the auction-level heterogeneity $\Xst$.
\end{ass}

In a first-price auction, the bidder with the highest bid wins and pays his own bid price. \citet{GPV2000} provides a one-to-one mapping between the conditional value distribution $f^{V|\Xst}(v|\xst)$ and the conditional bid distribution $f^{X|\Xst}(x|\xst)$ given that the competition $n$ is known. Thus, the identification of the conditional value distribution boils down to recover the conditional bid distribution $f^{X|\Xst}(x|\xst)$ from the bid data. If the data record all bids in each auction, the conditional independence property passes from values to bids. Consequently, the joint distribution of three independent bids, e.g., $X_1$, $X_2$, and $X_3$, denoted as $f(x, y, z)$, has the following multiplicatively separable structure:
\begin{eqnarray}
f(x, y, z) = \int_{\XXst}  \underbrace{f^{X|\Xst}(x|\xst) f^{X|\Xst}(y|\xst) f^{X|\Xst}(z|\xst)}_{\text{repeated measurements}} f^{\Xst}(\xst) d \xst,
\label{eq_joint_repeated}
\end{eqnarray}
based on which the conditional densities $f^{X|\Xst}(x|\xst)$ can be identified via eigenfunction decomposition \citep{hu2008instrumental}. The main idea is to exploit that the bids are repeated measurements of UH. Under Assumption \ref{cond_id}, their correlation reveals how UH affects the bids. Specifically, the observed joint distribution on the left-hand side of \eqref{eq_joint_repeated} identifies the conditional and marginal distributions on the right-hand side. 

Unfortunately, the auctioneer often does not record all bid information, and instead only records the most competitive bids. That is, the data essentially record a few order statistics of all bids, under which the conditional independence condition fails to hold.  This is because order statistics are ordered by definition. 

In an ascending auction, the bidder with the highest bid wins and pays the second highest submitted price, so a weakly dominant strategy is to continue bidding until the standing bid reaches one's own value. Therefore, all bidders bid their own values except the one with the highest value, who can simply outbid the second highest value by a small amount. That is, the highest bid and the second highest bid reveal essentially the same information regarding the second highest value, indicating that the highest bid is redundant. Because of this particular auction format, it is impossible to observe the highest value from the bids. Equivalently, we can view the auction as every one bids her/his value, but the auction fails to observe the highest bid/value.  Consequently, one cannot follow the aforementioned identification results to recover the conditional value distribution $f^{V|\Xst}(v|\xst)$, because the conditional independence condition fails. 
    
Facing the data limitation of incomplete bids, this paper focuses on identifying the conditional bid distribution $f^{X|\Xst}(x|\xst)$ for both first-price and ascending auctions from any three consecutive order statistics of all bids, i.e., $\{X_{r-2:n}, X_{r-1:n}, X_{r:n}\}$, where $X_{r-2:n}\le X_{r-1:n}\le X_{r:n}$. Once the conditional bid distribution is identified, the conditional value distribution can be identified using the one-to-one mapping between the bid and the value. 

Let  $\mathcal{V}$, $\mathcal{X}$, and $\XXst$ denote the supports of the distributions of the random variables $V$, $X$, and $\Xst$, respectively.  We first introduce the following regularity assumption. 
\begin{ass}
(Bound and Continuity)  \label{ass_bound} The joint density of $X$ and $\Xst$ admits a bounded and continuous density with respect to the product measure of some dominating measure $\mu$ (defined on $\mathcal{X}$) and the Lebesgue measure on  $\XXst$. All marginal and conditional densities are also bounded and positive. 
\end{ass}

We use $f_{r-2,r-1,r:n}(\cdot)$ and $f_{r-2,r-1,r:n}(\cdot|\xst)$ ($r \geq 3$) to represent the unconditional and conditional joint probability density functions (PDF) of the three order statistics, respectively, and $f^{X}_{r:s}(\cdot)$  and $f^{X \mid \Xst}_{r:s}(\cdot|\xst)$ represent the unconditional and conditional PDF of the $r$th order statistic of measurements ${X}$ out of a sample of size $s$ ($r \leq s$). 

The identification exploits the fact that the conditional joint distribution of three consecutive order statistics has a multiplicative separable structure. Specifically, the unconditional joint distribution, which can be estimated from the data, can be expressed as 
\begin{align}
\nonumber
&  \qquad f_{r-2,r-1,r:n}(x, y, z) =\int_{\XXst} f_{r-2,r-1,r:n}(x,y,z|\xst) f^{\Xst}(\xst) d \xst \\
& = c_{r,n} \underbrace{\mathbbm{1}(x \le  y \le z)}_{\text{correlation}} \int_{\XXst}  \underbrace{f^{X|\Xst}_{r-2:r-2}(x|\xst)  f^{X|\Xst}(y|\xst) f^{X|\Xst}_{1:n-r+1}(z|\xst)}_{\text{multiplicatively separable}} f^{\Xst}(\xst)d \xst,
\label{eq_joint1}
\end{align}
where $c_{r,n}=\frac{n!}{(r-2)! \cdot (n-r+1)!}$, and $\mathbbm{1}(\cdot)$ is the indicator function. The first equality holds by the law of total probability, and the second extends \citet{luo2020identification}'s Lemma 1 to three consecutive order statistics.\footnote{The joint distribution of any three order statistics does not have such a multiplicatively separable structure, i.e.,  $f_{r,s,t:n}(x,y,z)  \sim f(x)f(y)f(z) [F(x)]^{r-1} [F(y)-F(x)]^{s-r-1}[F(z)-F(y)]^{t-s-1}[1-F(z)]^{n-t}$, where $r<s<t$, see \cite{david2004order}. We derive Equation \eqref{eq_joint1} in the online supplement Section S.1.1.} This joint distribution of the consecutive order statistics has a semi-separable structure in the sense that we can separate the observed joint density function into the integration of three density functions, which is similar to \eqref{eq_joint_repeated} in the measurement error literature, but it has an extra restriction by the nature of order statistics, $\mathbbm{1}(x \le y \le z)$, which cannot be separated. This semi-separable structure precludes us from readily borrowing the same identification procedure in the existing literature to identify the conditional latent distributions directly. 

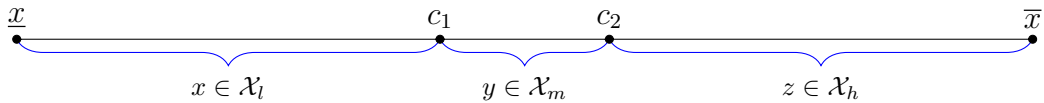
\begin{figure}[h!] 
\centering
\caption{Discretization.} \label{fig:discret}
\begin{tikzpicture}[scale=0.750]


%
%

\pgfkeys{/tikz/bracetemp/.code n args={4}{
\tikzmath{
coordinate \adjvset,\bracesp;
\adjvset=#4;
\bracesp1=#1;
\bracesp7=#2;
\bracesp4=($(\bracesp1)!0.5!(\bracesp7)+#3$);
\bracesp2=(\bracesp1)+(\adjvset);
\bracesp3=(\bracesp4)-(\adjvset);
\bracesp5=(\bracesp4)+(\adjvsetx,-\adjvsety);
\bracesp6=(\bracesp7)-(\adjvsetx,-\adjvsety);
}
}}

\draw [decorate,blue, decoration={brace,amplitude=10pt,mirror}](0,0) -- (7.5,0) node [black,midway,yshift=-0.65cm]{\footnotesize $x \in \mathcal{X}_l $};
\draw [decorate,blue, decoration={brace,amplitude=10pt,mirror}](7.5,0) -- (10.5,0) node [black,midway,yshift=-0.65cm]{\footnotesize $y \in \mathcal{X}_m$};
\draw [decorate,blue, decoration={brace,amplitude=10pt,mirror}](10.5,0) -- (18,0) node [black,midway,yshift=-0.65cm]{\footnotesize $z\in \mathcal{X}_h$};

\filldraw 
(0,0) circle (2pt) node[align=left,   above] {$\underline x$} --
(7.5,0) circle (2pt) node[align=center, above] {$c_1$}     -- 
(10.5,0) circle (2pt) node[align=right,  above] {$c_2$} --
(18,0) circle (2pt) node[align=right,  above] {$\overline x$};





\end{tikzpicture}
\end{figure}

Fortunately, the restriction by the indicator function can be safely circumvented if we divide the original support by two cutoff points $c_1$ and $c_2$, where $c_1 < c_2$, to separate the support into three parts, referred to as ``low," ``middle," and ``high," and denote them as $\mathcal{X}_l \equiv \{x: x \le c_1\}, \mathcal{X}_m \equiv[c_1, c_2]$, and $\mathcal{X}_h \equiv \{x: x \ge c_2\}$, respectively. Our context of three order statistics calls for three-part discretization, which extends \citet{luo2020identification}'s two-part discretization using two order statistics and an IV. The separable structure of the joint distribution $f_{r-2,r-1,r:n}(x, y,z)$ reappears if we always restrict $x \in \mathcal{X}_l$, $y\in \mathcal{X}_m$, and $z\in \mathcal{X}_h$. Specifically, if $x \in \mathcal{X}_l $, $y \in \mathcal{X}_m$, and $z \in \mathcal{X}_h$, the joint distribution can be expressed as
\begin{eqnarray}
f_{r-2,r-1,r:n}(x, y,z) =c_{r,n} \cdot  \int_{\XXst}  f^{X|\Xst}_{r-2:r-2}(x|\xst) f^{X|\Xst}(y|\xst) f^{X|\Xst}_{1:n-r+1}(z|\xst)  f^{\Xst}(\xst) d \xst, 
\label{eq-jointsep}
\end{eqnarray}
which has the same structure as the measurement error models but a different conceptual interpretation  for each component. Figure \ref{fig:discret} provides a visualization of the discretization. 

Following the identification strategy developed in \cite{hu2008instrumental}, we introduce the following integral operator that associates a function of two variables. 

\begin{definition} 

Let $L_{x|\tau}$ denote an operator that maps function $g$, where $g \in \mathcal{G}(\mathcal{T})$, to $L_{x|\tau}g \in \mathcal{G}(\mathcal{X}_l)$; and $H_{x|\tau}$ maps function $g$, where $g \in \mathcal{G}(\mathcal{X}_h) $, to $H_{x|\tau}g \in  \mathcal{G}(\mathcal{T})$. Specifically, the two operators are defined as
$$[L_{x|\tau}g](x) \equiv \int_{\mathcal{T}} f^{X \mid \Xst }(x|\tau) g(\tau) d\tau ~\quad\mbox{and}\quad~
[H_{x|\tau}g](\tau) \equiv \int_{\mathcal{X}_h} f^{X \mid \Xst }(x|\tau) g(x) dx.
$$
\end{definition}
Note that both operators involve a segment of bid support $\mA$. We further introduce another linear operator based on the joint distribution and the diagonal operator defined as follows. In particular, for a given $y \in \mathcal{X}_m$, let $J_y$ denote an operator mapping $g \in \mathcal{G} (\mathcal{X}_h)$ to 
$J_{y} g \in \mathcal{G}(\mathcal{X}_l)$:
\begin{equation*}
[J_{y}g](x) \equiv  \int_{\mathcal{X}_h} f_{r-2,r-1,r:n}(x, y, z) g(z)dz.
\end{equation*}
Given a particular partition $\{\mA_l, \mA_m, \mA_h\}$, $J_y$ is defined for every given $y$ in $\mA_m$. Let $\Delta_{X=y,\Xst}$ denote the diagonal operator mapping $g\in \mathcal{G}(\XXst)$ to $\Delta_{X=y,\Xst}g \in \mathcal{G}(\XXst)$:
\begin{equation*}
[\Delta_{X=y,\Xst}g](\xst) \equiv c_{r,n} f^{X \mid \Xst}(y|\xst)f^{\Xst}(\xst)g(\xst).
\end{equation*}

We derive the equivalence of operators in the online supplement Section S.1.2 as follows:
\begin{eqnarray}
J_{y} = L_{X_{r-2:r-2}|\Xst} \Delta_{X=y,\Xst}H_{X_{1:n-r+1}|\Xst},
\label{eq_joint3}
\end{eqnarray}
based on Equation \eqref{eq-jointsep} and by exploiting the following features: (i) an interchange of the order of integrations (justified by Fubini's theorem), (ii) the definition of $H_{X_{1:n-r+1}|\Xst}$, (iii) the definition of $\Delta_{X=y,\Xst}$ operating on $H_{X_{1:n-r+1}|\Xst}g$, and (iv) the definition of $L_{X_{r-2:r-2}|\Xst}$ operating on $[\Delta_{X=y,\Xst}H_{X_{1:n-r+1}|\Xst}g]$. Note that such equivalence between the operators holds for any value of $y \in \mathcal{X}_m$. 

For identification, we impose the following injective assumption.
\begin{ass}(Injective)  \label{ass_injective} There exists one division of the domain such that the operators $L_{\Xst|X_{r-2:r-2}}$ and $H_{X_{1:n-r+1}|\Xst}$ are injective for $\mathcal{G}=\mathcal{L}^2$, where $\mathcal{L}^2(\mathcal{X})$ denotes the set of all square integrable functions with domain $\mathcal{T}$ and $\mathcal{X}_h$, respectively.  
\end{ass}
An operator $A$ is injective if $Af = Ag$ implies $f = g$ for any $f, g$ in the domain of $A$. A linear operator being injective is equivalent to the family of kernel functions used to define the operator being complete; see \cite{hu2008instrumental}.
In our context, if the family of distributions $\{f^{X|\Xst}_{r-2:r-2}(x| \xst): x \in  \mathcal{X}_l \}$ is complete over $\mathcal{L}^2(\XXst)$, that is, the unique solution $\tilde{g}$ to the equation $\int_{\XXst}  g(\xst)f^{X|\Xst}_{r-2:r-2}(x| \xst) d \xst = 0$ for all $x \in  \mathcal{X}_l$ is $\tilde{g}(\cdot) = 0$, then $L_{\Xst|X_{r-2:r-2}}$ is injective under Assumption \ref{ass_bound}. We further provide conditions on the parental distributions under which the family of the order statistics' distributions is complete in the online supplement Section S.1.3. However, the equivalence between the injectiveness of operator $H_{X_{1:n-r+1}|\Xst}$ and the completeness of the kernel function family $\{f^{X|\Xst}_{1:n-r+1}(x| \xst):\xst \in \XXst \}$ over $\mathcal{L}^2(\mathcal{X}_h)$ is not straightforward, because the operator is defined only in a segment of the support. We prove that as long as the original distribution family is complete, i.e., $\{f^{X|\Xst}_{1:n-r+1}(x| \xst):\xst \in \XXst \}$ over $\mathcal{L}^2(\mathcal{X})$ is complete, there exists at least one division of the support such that operator $H_{X_{1:n-r+1}|\Xst}$ is injective. See the online supplement Section S.1.3. 



Completeness of the relevant family of distributions provides one way to characterize the injectivity of an operator.
Intuitively, the family of distributions  $\{f^{X | \Xst} (x |\tau): x \in \mathcal{X} \}$ being complete implies there is {\it sufficient variation} in the conditional density of $X$ across different values of $\Xst$. An example for such a complete distribution is a normal distribution with mean $\xst$ and variance 1. On the other hand, if the conditional density of $X$ does not vary sufficiently across $\tau$, such as the standard normal distribution, the distribution family is not complete. Obviously in such a scenario, $X$ is independent of $\Xst$, and hence we can easily find $g \neq 0$ such that $\int g(\tau) f^{X | \Xst}(x | \tau) d \tau = 0$ for any $x$.

Assumption \ref{ass_injective} also specifies that we consider the identification with $\mG =\mL^2$. Such consideration is due to the following two reasons. First, this space is sufficiently large such that the density can be sampled everywhere, which ensures a one-to-one mapping between a density function and its corresponding operator. Thus, the density function can be uniquely determined by the associated operator with such a choice of $\mathcal{G}$.\footnote{The space $\mathcal{G}=\mathcal{L}^2$ is sufficiently rich, because 
$
f^{X}_{r-2:r-2}(x|\xst_0) = \lim\limits_{n \rightarrow \infty} [L_{X_{r-2:r-2}|\Xst}g_{n, \xst_0}](x),$
where $g_{n, \xst_0}(\xst) = n\mathbbm{1}(|\xst - \xst_0| \leq n^{-1} )$, a sequence of bounded and square-integrable functions. } Second, it is a Hilbert space if equipped with the norm $\|g\|_{\mL^2} = \left(\int_{\mA} g^2(x) dx\right)^{1/2}$ for any $g \in \mG(\mA)$. One advantage of considering Hilbert spaces is that it is easier to use properties of the operators such as $L_{X_{r-2:r-2}|\Xst}$ and $H_{X_{1:n-r+1}|\Xst}$ later, because there are many existing theoretical results developed for operators defined in Hilbert spaces. For instance, it is straightforward to define the adjoint operator by using the concept of inner product in Hilbert spaces. It is also worth noting that this space is smaller than the $\mathcal{L}^1$ space adopted in \cite{hu2008instrumental}, 
which is a Banach space.  

If an operator is injective, its inverse is well-defined, but may be defined over a restricted domain. We further prove that $L_{X_{r-2:r-2}|\Xst}$ is surjective in addition to being injective, so that the domain of its inverse is the whole space $\mL^2(\mA)$. This is important for proving the equivalence of operators defined in the data and in the distributions to be identified. We summarize this result in the following lemma and relegate the proof to the online supplement Section S.1.4.

\begin{lemma}
If  Assumptions \ref{cond_id}-\ref{ass_injective} hold,  then $L^{-1}_{X_{r-2:r-2}|\Xst}$ exists and is densely defined over $\mL^2(\mA_l)$.
\label{lema_1}
\label{le-bijective}
\end{lemma}
Lemma \ref{le-bijective} essentially indicates that operator $L_{X_{r-2:r-2}|\Xst}$ is surjective if it is injective. We use the following simple example to facilitate understanding the necessity of the surjective property and the difference between linear operators and matrices. Suppose that $\mathcal{D}^1$ and $\mathcal{D}^2$ are two linear spaces, and $L$ is a linear transformation from $\mathcal{D}^1$ to $\mathcal{D}^2$. If both $\mathcal{D}^1$ and $\mathcal{D}^2$ are finite-dimensional, $L$ is injective if and only if it is surjective. In particular, if dim($\mathcal{D}^1$) = dim($\mathcal{D}^2$) and $L$ is associated with a square matrix $A$, then $L$ is both injective and surjective if and only if $A$ has full rank. But this relationship does not trivially hold in infinite-dimensional cases. For example, let $\{e_i\}^\infty_{i=1}$ be the basis of $\mathcal{D}^1$ as well as $\mathcal{D}^2$. We assume that $Le_i = e_{i+1}$ for every $i \ge 1$. Such an operator $L$ is obviously injective but not surjective, {because the base $e_1$ is missing in its range.}

Since $H_{X_{1:n-r+1}|\Xst}$ is injective under Assumption \ref{ass_injective},
we can eliminate the common operator $H_{X_{1:n-r+1}|\Xst}$  by equivalence of operators specified in Equation \eqref{eq_joint3} for any two different values of $y$, i.e., $y_1$ and $y_2$, leading to the following main equation for identification:
\begin{eqnarray}
J_{y_1} J^{-1}_{y_2}&=& L_{X_{r-2:r-2}|\Xst} \Delta_{X=y_1,\Xst} \Delta^{-1}_{X=y_2,\Xst}L^{-1}_{X_{r-2:r-2}|\Xst}. \label{eq_joint8}
\end{eqnarray}
By Lemma \ref{le-bijective}, the relation \eqref{eq_joint8} is established over a dense subset of $\mathcal L ^2(\mathcal X_l)$. In fact, it can be further  extended to the full space $\mathcal L ^2(\mathcal X_l)$ by leveraging the extension procedure of linear operators. 
This equation ensures that operator $J_{y_1} J^{-1}_{y_2}$ can be represented as an eigenvalue-eigenfunction decomposition with the two unknown operators $L_{X_{r-2:r-2}|\Xst}$ and $\Delta_{X=y_1,\Xst}\Delta^{-1}_{X=y_2,\Xst} $ being the eigenfunctions and eigenvalues, respectively. Consequently, diagonalizing operator $J_{y_1} J^{-1}_{y_2}$, which can be computed from the data directly since it is defined using observable densities, provides the eigenfunctions $L_{X_{r-2:r-2}|\Xst}$, indexed by the latent UH, and further provides the unobserved densities of order statistic $X_{r-2:r-2}|\Xst$. 

Note that there are three features prevalent in identification using decomposition: The identification may not be unique; the identification is up to scales; the identification is up to ordering and location. We tackle the three issues one at a time below.

\subsection*{Unique Decomposition}

To guarantee unique decomposition, we impose restrictions on the relationship between observed measurement $X$ and UH $\Xst$ in segment $\mathcal{X}_m$.  
\begin{ass}(Distinct)  \label{ass_distinct} there exists one division of the domain such that, for all $\xst_1, \xst_2 \in \XXst$, the set $\{(y_1,y_2):  \frac{f^{X|\Xst}(y_1|\xst_1)}{f^{X|\Xst}(y_2|\xst_1)} \neq  \frac{f^{X|\Xst}(y_1|\xst_2)}{f^{X|\Xst}(y_2|\xst_2)}, where ~ (y_1,y_2) \in  \mathcal{X}_m \times \mA_m\}$ has positive probability whenever $\xst_1 \neq \xst_2$.
\end{ass}
This assumption is weaker than assuming that the associated operator is injective in segment $\mathcal{X}_m$. Note that we just need one division where such an assumption holds. This assumption fails only if the distribution of the measurement conditional on the latent factor is the same at the two distinct values $\xst_1$ and $\xst_2$.

Assumption \ref{ass_distinct} guarantees unique eigenvalues, so that conducting the decomposition to operator $J_{y_1} J^{-1}_{y_2}$ identifies operator $L_{X_{r-2:r-2}|\Xst}$, and thus identifies the conditional density $f^{X|\Xst}_{r-2:r-2}(x|\xst)$, for $x \in \mathcal{X}_l$. However, such identification is up to scales. That is,  the conditional density $f^{X|\Xst}_{r-2:r-2}(x|\xst)$ is identified as the true density multiplied by an unknown constant, which could differ for each UH. The existing literature relies on the property that the total probability is equal to 1 for each conditional distribution to pin down the scales. Such an approach is not feasible in our framework because, from the decomposition, we only identify the conditional distribution in one segment of the full support, i.e., $\mathcal{X}_l$. Mover, one can neither pin down the ordering or the actual values of UH, which calls for extra restrictions. 
To proceed, we propose to leave the ordering of the UH  and the scales in the low segment as undetermined and proceed to identify the conditional distributions in the other two segments first. In this procedure we mainly use Equation \eqref{eq_joint8}. One main feature worth noting during this process is that we keep the value of the UH consistently matched across the three segments. Furthermore, these scales are the same for the same UH in the same segment but may vary across UH or segments. Given these, we can then pin down the scales and ordering in what follows.

\subsection*{Unique Scale}

Note that we can identify the conditional distributions in all three segments up to different scales. That is, each segment of the conditional distribution is associated with one scale parameter, so together there are three scale parameters to pin down for each conditional distribution. These scales can then be pinned down by invoking the continuity of the component PDFs and the total probability argument. First, the PDFs identified separately in the three segments should be the same at the cutoff points  due to the continuity of the true conditional distributions. Second, the fact that each conditional distribution should integrate to 1 provides the third restriction on the scales. These restrictions uniquely identify the scales. 

\subsection*{Unique Ordering and Location}


Given that the conditional distributions are identified in the full support, we provide a condition using the auction setting to pin down the exact location of the UH. Specifically, letting UH be the unobserved quality of the auctioned item, we would expect that bidders' values/bids are, on average, higher and of better quality. For instance, in second-hand automobile auctions, omitted details from the car description, such as 
dents and scratches, are revealed upon pre-auction inspection and enter bidder values. 

 \begin{ass}(Monotonicity and Location) \label{ass_order} The expected value/bid is strictly monotone with UH; that is, $E(X|\Xst=\xst)$ is strictly monotone with $\xst$ for all $\xst \in \XXst$. Moreover, we assume that the support of UH is [0, 1]. 
\end{ass}
The monotonicity assumption is useful to pin down UH's relative ordering. However, its exact location/value is still unidentified. That is, one could always apply a monotone transformation to the UH and obtain an observationally equivalent model that satisfies all assumptions. To pin down UH's exact location, we normalize its support to be $[0,1]$, which is without loss of generality. Such a normalization is similar to the mean zero normalization.



\begin{theorem}
If  Assumptions \ref{cond_id}-\ref{ass_order} are satisfied, conditional bid distribution $f^{X|\Xst}(x|\xst)$ for $x\in \mathcal{X}$ and $\xst \in \XXst$ and UH's distribution $f^{\Xst}(\xst)$ for any $\xst \in \XXst$ are identified using any three consecutive order statistics of bids.
\label{th-identification}
\end{theorem}
We summarize the main steps of the proofs below and leave the details to the online supplement Section S.1.6.\footnote{We thank Yingyao Hu and Ji-Liang Shiu for valuable insights about proving the theorem.} First, we identify operator $L_{X_{r-2:r-2}|\Xst}$ from the decomposition of Equation (\ref{eq_joint8}). Such identification is unique by Assumption \ref{ass_distinct}, but up to scales and location. Second, we identify the operator $H_{X_{1:n-r+1}|\Xst}$ up to different scales, similar to the identification of $L_{X_{r-2:r-2}|\Xst}$. Third, for any value $y\in  \mathcal{X}_m$, we can identify operator $\Delta_{X=y,\Xst}$ up to the same scales for all $y$ once we plug the identified operators $L_{X_{r-2:r-2}|\Xst}$ and $H_{X_{1:n-r+1}|\Xst}$ into Equation \eqref{eq_joint3}. Using the one-to-one mapping between operators and the associated densities, we then identify the unobserved densities $f^{X|\Xst}_{r-2:r-2}(x|\xst)$ for $x \in \mathcal{X}_l$, $f^{X|\Xst}(y|\xst) f^{\Xst}(\xst)$ for $y\in  \mathcal{X}_m$, and $f^{X|\Xst}_{1:n-r+1}(z|\xst)$ for $z \in  \mathcal{X}_h$ up to scales. The scales are the same in the same segment but may vary across different segments. Furthermore, we show that the one-to-one mapping between the distribution of an order statistic and its parent distribution can be extended from the full support to a segment. Thus, we identify the conditional distribution up to different scales in all three segments. Lastly, the scales are then pinned down using three restrictions. 

Once the conditional bid distributions are identified as in Theorem \ref{th-identification}, we can exploit the one-to-one mapping between the conditional value and bid distributions to recover the conditional value distributions, which are the target of interest. Specifically, for ascending auctions, where bidders' weakly dominant strategy is to bid their values, the conditional value distribution is the same as the conditional bid distribution;\footnote{Many empirical studies adopt the same assumption in ascending auctions; see, e.g., \citet{lu2008estimating}, \citet{aradillas2013identification}, and \citet{hortaccsu2021empirical}. We exclude other possible bidding strategies such as jump bidding allowed in \cite{haile2003inference}. Such abstraction is a good approximation for online auctions and button auctions. For instance, eBay allows bidders to set up a proxy bid.} for first-price auctions, we can identify the conditional value distribution by exploiting the one-to-one mapping established in \cite{GPV2000}. We summarize this result in the following Corollary.

\begin{corollary} \label{corollary1} If  Assumptions \ref{cond_id}-\ref{ass_order} are satisfied, the conditional value distribution \\
$f^{v|\Xst}(v|\xst)$ for $v \in \mathcal{V}$ and $\xst \in \XXst$ and the latent variable's distribution $f^{\Xst}(\xst)$ for $\xst \in \XXst$ are identified using any three consecutive order statistics of bids.
\label{coro0}
\end{corollary}

The identification results in Theorem \ref{th-identification} are achieved under the assumption that the reserve price is not binding. However, in practice, the reserve price appears to be binding in many cases, leading to a truncation in the observed bid distribution. We show in the following corollary that we can still identify the bid/value distribution with a truncation. We can also identify the conditional probability of the truncation when the number of potential bidders is observed. 

\subsection*{Reserve Price for Ascending Auctions}
If the reserve price is binding, the optimal bidding strategy for any bidder is to submit the optimal bid computed without reserve prices when such an optimal bid is above the reserve price, and to not bid otherwise. Therefore, the presence of a binding reserve price $R$ creates a truncation in the observed bid distribution, i.e., $\tilde F^{X|\Xst}(x|\xst) \equiv \frac{F^{X|\Xst}(x|\xst)-F^{X|\Xst}(R|\xst)}{1-F^{X|\Xst}(R|\xst)}$, where $ x\in [R,\overline{x}]$. Let $n$ denote the number of actual bidders and $N$ denote the number of potential bidders. In first-price auctions, even if entry is exogenous, the observed bid distribution depends on both $N$ and $n$, while in ascending auctions, it only depends on $n$. Therefore, to illustrate the intuition, we focus on ascending auctions.

Under such a situation, even with a truncation caused by a binding reserve price, we can still follow the identification strategy in Theorem \ref{th-identification} to identify the truncated CDF $\tilde F^{X|\Xst}(x|\xst)$, PDF $\tilde f^{X|\Xst}(x|\xst)$, and the marginal distribution of the UH without information on $N$ as long as $n$ is known. Specifically, the joint distribution of three consecutive active bids with a bidding reserve price can be expressed as
\begin{eqnarray*}
\tilde{f}_{r-2,r-1,r:n}(x, y, z) = c_{r,n} \cdot \mathbbm{1}(x \!\le\! y \!\le\! z) \cdot  \int_{\XXst}  \tilde f^{X|\Xst}_{r-2:r-2}(x|\xst)  \tilde f^{X|\Xst}(y|\xst) \tilde f^{X|\Xst}_{1:n-r+1}(z|\xst) f^{\Xst}(\xst)d \xst.
\end{eqnarray*}
A few features are worth noticing. First,  identification using eigen-decomposition applies regardless of whether $N$ is observed, as the bidding strategy does not vary with $N$ under exogenous entry. Second, without observing bids below the reserve price, there is no information to identify the bid/value distribution for this segment. Lastly, we establish that we can identify the conditional probability of the truncation $F^{X|\Xst}(R|\xst)$. 

\begin{corollary} In ascending auctions, when $N$ is observed and has a large support, the conditional probability of truncation $F^{X|\Xst}(R|\xst)$ is identified using the distribution of the number of actual bidders conditional on the potential bidders. Therefore, for all $x\geq R$, $F^{X|\Xst}(x|\xst)$ is identified from $\tilde F^{X|\Xst}(x|\xst) \equiv \frac{F^{X|\Xst}(x|\xst)-F^{X|\Xst}(R|\xst)}{1-F^{X|\Xst}(R|\xst)}$. 
\label{coro1}
\end{corollary}
The detailed proof for Corollary \ref{coro1} can be found in the online supplement Section S.1.7. Intuitively, the distribution of $n$ conditional on $N$ is a mixture of binomial distributions with the success probability being the conditional truncated probability. That is,
\begin{eqnarray}
\Pr(n|N) & = \int_{\xst \in \XXst} C_{N,n} [1-F^{X|\Xst}(R|\xst)]^n [F^{X|\Xst}(R|\xst)]^{N-n}d F^{\Xst}(\xst),
\end{eqnarray}
where $\Pr(n|N)$ is estimable from the data, $C_{N,n}$ is a constant,  $F^{\Xst}(\xst)$ can be treated as known, and conditional truncation probability $F^{X|\Xst}(R|\xst)$ is the object of interest. This  is similar in structure to but differs conceptually  from the identification in the mixture literature \citep{gut2005probability}, where the goal is to identify the mixture distribution with the success probability taking any value in $[0,1]$. We show that our identification problem can be viewed as the dual problem of that by changing variables in the integral. 

\subsection*{The Number of Order Statistics}

Our discussion so far assumes that three consecutive order statistics of bids are available. There are various ways to extend this main identification result. First, the required number of consecutive order statistics reduces to two if there exists an instrument that is independent of the bids conditional on  UH; see \citet{luo2020identification}.\footnote{Measurement error approaches are inapplicable when only one order statistic, such as the winning bid, is observed. This calls for alternative strategies, such as density discontinuity approaches first proposed by \citet{GL2018}.} Second, while consecutiveness barely restricts the data with incomplete bids, exploiting the Markov property of order statistics relaxes this requirement. In the online supplement Section S.3, we show that any four order statistics identify the model.\footnote{The idea of using Markov property for dealing with UH and incomplete bid data simultaneously is first explored in \cite{M2017}, who uses five order statistics in finite UH framework.}

\section{Sieve Maximum Likelihood Estimation} \label{sec:sieve}

Note that conducting counterfactual policy analysis requires one to estimate the joint distribution of UH and bidder private values. In principal, the conditional bid distribution and UH's  marginal distribution could be estimated fully nonparametrically by following the constructive identification argument step-by-step. Specifically, one could do a partition in the full support and conduct eigenfunction decomposition to estimate the distribution of the order statistics in the three segments,  then use the one-to-one mapping between the distribution of an order statistic and its parent distribution to estimate the parent distribution. Such a fully nonparametric estimator not only poses a high demand on the data but is also of low efficiency, as it depends critically on the partition of the support and involves sequential estimation. 

Considering the fact that, in applications, the analyst oftentimes can only access modest-sized data, we propose to estimate these two densities using the method of sieves \citep{grenander1981abstract, shen1997methods, chen1998sieve, chen2007large} to fully exploit variations in the data instead of relying on a particular partition. We establish consistency and convergence rates for such estimators. 



Our strategy is to first provide some regularity assumptions on the sieve approximation for consistency, which usually depends on the smoothness of the function to be approximated and the complexity of the sieve space. Such complexity is characterized by its upper bound and bracketing numbers.\footnote{In contrast, \cite{hu2008instrumental} uses a covering number to characterize complexity.} To further understand the scope of our general results, the online supplement Section S.2.2 proves that the sieve space constructed by either B-spline or Bernstein basis functions, which are popular sieve spaces in auctions, satisfies the regularity assumptions, and thus, the estimator is consistent. 



We represent the log likelihood function of the joint distribution of the three consecutive order statistics, i.e., $\text{data}\equiv\{X_{r-2:n}=x^i, X_{r-1:n}=y^i, X_{r:n}=z^i\}^m_{i=1}$, as follows:
\begin{eqnarray}
\log L(\mbox{data}; f^{X|\Xst}, &f^{\Xst}) =\frac{1}{m}  \frac{n!}{(r-3)!(n-r)!} \sum^m_{i=1} \log \int_{\xst}  [F^{X|\Xst}(x^i|\xst)]^{r-3} f^{X|\Xst}(x^i|\xst) \notag\\
& f^{X|\Xst}(y^i|\xst)[1-F^{X|\Xst}(z^i|\xst)]^{n-r}f^{X|\Xst}(z^i|\xst)  f^{\Xst}(\xst)d \xst.
\end{eqnarray}


As both the conditional density and the marginal density can be derived from a joint density, we propose to approximate joint distribution $f^{X, \Xst}(x,\xst)$ by using tensor product bases of univariate series. Specifically, let $\mB_m$ be the finite-dimensional sieve space and $\xi_1, \ldots, \xi_{p_m}$ be its basis, where $p_m$ is the number of basis functions in the sieve space.


With slight abuse of notation, we denote the sieve representation of this joint distribution as $\mathfrak{f}$. We then represent the marginal distribution, the conditional distribution, and CDF of such a conditional distribution as follows:
\begin{eqnarray}
f^{\Xst}(\xst)&=&\int_{\mathcal{X}}f^{X, \Xst}(x,\xst) d x \simeq \int_{\mathcal X}\mathfrak{f}(x,\xst) d x \label{marginal},\\
f^{X|\Xst}(x|\xst)&=&\frac{f^{X, \Xst}(x,\xst)}{f^{\Xst}(\xst)} \simeq \frac{ \mathfrak{f}(x,\xst)}{\int_{\mathcal X}\mathfrak{f}(x,\xst) d x} \label{conditional},\\
F^{X|\Xst}(x|\xst)&=& \int^x_{-\infty}  f^{X|\Xst}(t|\xst)d t  \simeq  \frac{ \int^x_{-\infty} \mathfrak{f}(t,\xst) dt}{\int_{\mathcal X} \mathfrak{f}(x,\xst) d x}. \notag
\end{eqnarray}
Consequently, the sieve estimator for the joint distribution of the three observed consecutive bids can be represented as
\begin{eqnarray} \label{sieveMLE}
\hat{\mathfrak{f}} &=& \argmax_{\mathfrak{f} \in \mB_m}\log L\left(\mbox{data}; \frac{ \mathfrak{f}(x,\xst)}{\int_{\mathcal X} \mathfrak{f}(x,\xst) d x}, \int_{\mathcal X} \mathfrak{f}(x,\xst) d x \right).
\end{eqnarray}


Next, we show that under some regularity conditions the proposed sieve estimator for the joint distribution in Equation \eqref{sieveMLE} is consistent. Once the joint distribution is consistently estimated, the conditional and  marginal distributions, specified in Equations \eqref{marginal} and \eqref{conditional} respectively, are also consistently estimated. Let $ f_0^{X, \Xst}(x, \xst)$ denote the true joint density, and  let $ f_0^{X | \Xst}(x | \xst)$ and $f_0^{\Xst}(\xst)$ denote the true conditional density of $X$ given $\Xst = \xst$ and the marginal density of the latent variable, respectively. We introduce some regularity conditions.

\begin{ass}(Compactness)  \label{ass_comp} 
	$X$ has a compact support. Without loss of generality, we assume that its support is [0, 1]. 
\end{ass}

This compact support assumption is standard in the auction literature. Moreover, we can linearly transform random variables with compact support to ones that have support on $[0,1]$. Note that such a transformation has to be linear, rather than an arbitrary monotone transformation. The linear transformation is for convenience of using the observed data in estimation. The support of the two random variables, $X$ and $\Xst$, plays an important role in choosing an appropriate sieve space to perform maximum likelihood estimation. For example, the trigonometric sieve is inapplicable when the support is $\Re$. In this case, Hermite polynomials and B-splines are preferable. B-spline approximation is also useful when the support is compact. It is worth emphasizing that our identification results hold regardless of this normalization. 

\begin{ass}(Sieve approximation)  \label{ass_appro} 
	There exists $f_m^{X, \Xst}(x , \xst)$, which is represented in terms of the bases $\xi_1, \ldots, \xi_{p_m}$ in the sieve space, for some $\beta > 0$, such that
	\begin{align*}
		\|f_m ^{X, \Xst}(x, \xst) - f_0^{X, \Xst} (x, \xst)\|_{L_{\infty}([0, 1]^2)} = O(p_m^{-\beta}).
	\end{align*} 
\end{ass}
Assumption \ref{ass_appro} ensures that the joint density can be approximated sufficiently well in the sieve space. Consequently, by Equations \eqref{marginal} and \eqref{conditional}, both the conditional density and the marginal density can be approximately sufficiently well by functions in the sieve space. That is, with Assumption \ref{ass_appro}, there exist $f_m^{X | \Xst}(x | \xst)$ and $f_m^{\Xst}(\xst)$, both represented in terms of $\xi_1, \ldots, \xi_{p_m}$ in the sieve space, such that
\begin{align*}
	\|f_m^{X | \Xst}(x | \xst) - f_0^{X | \Xst}(x | \xst)\|_{L_{\infty}([0, 1]^2)} = O(p_m^{-\beta}), ~\mbox{and}~~
	\|f_m ^{\Xst}(\xst) - f_0^{\Xst} (\xst)\|_{L_{\infty}([0, 1]^2)} = O(p_m^{-\beta}).
\end{align*}


To study the asymptomatic properties of the proposed estimator, we first establish the relationship among the sieve estimator, the sieve representation, and the underlying true densities. Let $G(x, y, z; f^{X | \Xst}, f^{\Xst})$ be the log-likelihood function from one single observation that depends on the conditional density of $X$ given $\Xst = \xst$ and the marginal density of $\Xst$.
\begin{lemma} \label{lema_inequa}
	Let $\hat{f}_m^{X | \Xst} (x | \xst)$ and $\hat{f}_m^{\Xst} (\xst)$ denote the estimated conditional density of $X$ and the marginal density of the latent variable $\Xst$, respectively.  We have
	\begin{align}
	\nonumber
	\frac{1}{\sqrt m} \bG_m\left[\log \frac{G(x, y, z; \hat{f}_m^{X | \Xst}, \hat{f}_m^{\Xst})}{G(x, y, z; f_m^{X | \Xst}, f_m^{\Xst})}  \right] & \ge  \bP\left[\log \frac{G(x, y, z; f_m^{X | \Xst}, f_m^{\Xst})}{G(x, y, z; f_0^{X | \Xst}, f_0^{\Xst})}  \right] \\
	& ~~~
+ \bP\left[\log \frac{G(x, y, z; f_0^{X | \Xst}, f_0^{\Xst})}{G(x, y, z; \hat{f}_m^{X | \Xst}, \hat{f}_m^{\Xst})}  \right],
\label{eq_lemm2}
	\end{align}
	where $\bG_m = \sqrt{m}(\bP_m - \bP)$, $\bP_m$ denotes the empirical measure of data $(x_i, y_i, z_i)_{i = 1}^m$, and $\bP$ denotes the true distribution.
\end{lemma}

Lemma \ref{lema_inequa} holds by definition of $f_0$ and $\hat{f}$. The proof can be found in the online supplement Section S.2.1. To show consistency of the sieve estimator, we need to bind the left-hand side of Equation \eqref{eq_lemm2}. 
To accomplish this, we resort to empirical process theories and impose restrictions on the complexity of the sieve space. We first introduce the following two assumptions to characterize its complexity.

\begin{ass} [Bound of sieve space]
	\label{ass:bound}
The logarithm of the upper bound over $\mB_m$, denoted by $Q_m$, satisfies $\log\{\sup_{\mathfrak{f} \in \mB_m} \|\mathfrak{f}\|_{L_{\infty}([0, 1]^2)} \} \leq Q_m = O(\log \log m)$.
\end{ass}

\begin{ass} [Bracketing number] \label{ass-bracket}
	The $\epsilon$ bracketing number of the sieve space $\mB_m$ is of  order $O\left((e^{2Q_m}/\epsilon)^{p_m + 2}\right)$ for some constant $p_m = O(m^{\alpha})$ with $0 < \alpha < 1/2$. 
\end{ass}
Intuitively, $Q_m$ would be larger  for a larger space. We define the bracketing number following \cite{van199}. Specifically, given two functions $l$ and $u$, the bracket $[l ,u]$ is the set of all functions $f$ with $l \leq f \leq u$. An $\epsilon$-bracket is a bracket $[l, u]$ with $\|u - l\| \leq \epsilon$ under a certain norm $\|\cdot\|$. The $\epsilon$ bracketing number $N_{[]}(\epsilon, \mathcal{B}, \|\cdot\|)$ is the minimum number of $\epsilon$-brackets needed to cover $\mathcal{B}$. A larger $\epsilon$ bracketing number corresponds to a more complex sieve space. 

To guarantee consistency, we consider the function class $\mF_m$, defined by
\begin{eqnarray*}
	\left\{\log \frac{G(x, y, z; \tilde{f}_m^{X | \Xst}, \tilde{f}_m^{\Xst})}{G(x, y, z; f_{m}^{X | \Xst}, f_{m}^{\Xst})}:  \tilde{f}_m^{X | \Xst} = \frac{ \mathfrak{f}_m(x,\xst)}{\int_{\mathcal X} \mathfrak{f}_m(x,\xst) d x},
	\tilde{f}_m^{\Xst} = \int_{\mathcal X} \mathfrak{f}_m(x,\xst) d x \right\},
\end{eqnarray*}
where $\mathfrak{f}_m$ is represented in terms of $\xi_1, \ldots, \xi_{p_m}$ in sieve space $\mB_m$. If the complexity of sieve space $\mathcal B_m$ satisfies Assumptions \ref{ass:bound}-\ref{ass-bracket}, we are able to quantify the upper bound on $\mF_m$, which is the upper bound on the left-hand side of Equation \eqref{eq_lemm2}.

We now establish consistency of the proposed sieve estimator. 
\begin{theorem} \label{consistent}
	Under Assumptions \ref{ass_comp}-\ref{ass-bracket}, the proposed sieve MLE for the joint distribution is consistent. Moreover, both the conditional and marginal distributions are consistently estimated. That is,  
	\begin{align*} 
		\|\hat{f}_m^{X | \Xst} (x | \xst) - f_0^{X | \Xst}(x | \xst)\|_{L_2} \overset{p}{{\longrightarrow}} 0, & \text{ and }~ 
		\|\hat{f}_m^{\Xst} (\xst) - f_0^{\Xst}(\xst)\|_{L_2} \overset{p}{{\longrightarrow}} 0.
	\end{align*}
The convergence rate for these estimators is derived to be $B(m, p_m, Q_m)^{1/2}$, where\\
$B(m, K_m, Q_m) = e^{c_2Q_m} p_m \log p_m / \sqrt{m} + e^{c_2Q_m} / p_m^{\beta}$, with $c_2$ being a constant.
\end{theorem}





The detailed proof is given in the online supplement Section S.2.1. Note that we consider $L_2$ convergence of our proposed estimator. Establishing the (uniform) convergence rate is beyond the scope of this paper and thus left for future research. As pointed out in \cite{menzel}, the uniform rate depends on $r$, and the nonparametric MLE of the parent distribution obtained using order statistics may have a slower convergence rate near the tail of the parent distribution. The primary reason for the latter is that the mapping from the distribution of order statistics to the corresponding parent distribution may not be Lipschitz continuous. The derivative of this mapping may diverge near the tail. In this context, we found similar issues with respect to the proposed sieve MLE using Berstein polynomials from simulation studies. See the online supplement Section S.2.3.

\paragraph{The Conditional Value Distributions }Theorem \ref{consistent} concerns the distribution of UH and the conditional bid distributions. While the bid equals the value in ascending auctions, recovering the value distributions in first-price auctions requires several additional steps. First, we estimate the conditional bid quantile functions $\widehat{b}(\alpha|\xst)$ by inverting the estimated conditional bid distribution. That is, $\widehat{b}(\alpha|\xst)=\widehat{F}^{-1}(\alpha|\xst)$, where we have omitted  supscript $X|\Xst$ for simplicity. Second, following \citet{GPV2000}, we can recover the conditional value quantile function
\begin{eqnarray}
\widehat{v}(\alpha|\xst)=\widehat{b}(\alpha|\xst)+\frac{1}{n-1}\alpha\widehat{b}^{\prime}(\alpha|\xst), 
\end{eqnarray}
which allows constructing the conditional value density and distribution. By the continuous mapping theorem \citep{chung2001course}, the estimated conditional value quantile function, density, and distribution are also consistent. Moreover, if we impose higher-order smoothness assumptions on the value distribution, we may achieve a faster convergence rate, which is similar to the results in \cite{menzel}.

\section{Empirical Application} \label{sec:app}

In this section, we apply our methodology to an empirical analysis of judicial auctions in China. Chinese courts began holding online auctions in 2012 through taobao.com, the shopping site of Chinese e-commerce giant Alibaba. As of 2022, almost all of China's courts have registered on this judicial sales platform, auctioning assets ranging from cars, diamonds, property, land use rights, and Boeing 747s to company shares. As of December 2019, over 500,000 items have been sold, with turnover reaching about 1.3 trillion yuan on the Taobao judicial sales platform.\footnote{\href{https://global.chinadaily.com.cn/a/201912/26/WS5e0411dda310cf3e35580b5c_2.html}{Source: China Daily.}}

The court first posts the property-related information on taobao.com, including the appraisal value, obtained through a third-party appraisal company, and a starting price. Potential buyers can view the information page online and visit the property physically before the auction starts. Interested bidders can register to participate in the bidding by paying a security deposit and then bid in an ascending fashion. They can also set up automatic bidding.\footnote{On average, a sold item receives 55 bids from 3 bidders, suggesting that jump bidding may not be a big concern.} The highest bidder wins the object and pays his/her bid. 

\subsection{Data}

We collect a sample of residential property auctions from taobao.com, which contains all sales by the court in Jiangmen city of Guangdong Province between January 2018 and June 2020. We drop a few sales that are below ten thousand RMB or above five million RMB. In total, we have 477 auctions with 329 successful sales. By default, this court uses 70\% of the appraisal value as the starting price, which also serves as a reserve price. 

These auctions are subject to UH for many reasons. A third party provides appraisal based on available information at hand but may miss important details that become revealed upon careful study of the listing and a physical visit. For example, any unpaid electricity bills or property management fees of a sold property are the responsibility of the winning bidder. Some condos may have defects that are unknown to the appraisal firm. These unobserved factors constitute a significant portion of potential bidders' values. But how they enter bidder value is unknown. Therefore, it is preferable to retain flexibility when specifying how bidder value depends on UH and private information. 

Following the literature, we homogenize the bids by dividing them by the appraisal value.\footnote{For the homogenization to be valid, we need either 1) the appraisal value to be realized  before the realization of UH or 2) the seller or the third-party appraisal company to have the same access to UH but choose to ignore the additional knowledge.} We further rescale the homogenized bids by dividing them by the maximum value in estimation but report the results in homogenized terms for convenience. 
As usual, the highest and second-highest bids are close to each other, both revealing information about the second-highest value among all bidders.  To avoid redundant information, we use the highest bid as the second-highest value among all the bidders and exclude the second-highest bid from the data.\footnote{We obtained almost identical estimation and counterfactual results using the second highest bid as the second highest valuation.} 

Table \ref{sumstat} provides some summary statistics of our data. On average, each property is worth one million RMB, which is approximately \$140,000 USD. Only about $70\%$ of listings are sold successfully, at a transaction price close to the appraisal value on average.  

\begin{table}[ht]
\centering
\caption{Summary Statistics} \label{sumstat}
\begin{tabular}{clllll} \hline \hline 
Variable     & Obs & Mean     & Std. Dev. & Min      & Max      \\ \hline
appraisal value (million RMB) & 477 & 1.02 & 0.91  & 0.108 & 4.97 \\
\# of potential bidders & 477 & 5.285 & 5.867 & 0 & 31 \\
\# of bidders & 477 & 2.182    & 1.946     & 0        & 10       \\
sold         & 477 & 0.690    & 0.463     & 0        & 1        \\
$\frac{\text{winning bid}}{\text{appraisal value}}$  & 329 & 0.995    & 0.267     & 0.700      & 2.359    \\ \hline
\end{tabular}
\end{table}

\subsection{Empirical Model with a Binding Reserve Price}

Our empirical model accounts for the binding reserve price. Upon arrival, $N$ potential bidders observe the realization of UH $\xst$ and draw i.i.d. private values from $F^{X|\Xst}(\cdot | \xst)$. Those with a valuation higher than reserve price $R$ submit a bid equal to their value. As a result, the amount of truncation for a given UH is $F^{X|\Xst}(R|\xst)$, where $R=0.7$. 

Conditional on the number of potential bidders $N$, the probability of observing the bid vector $\boldsymbol{b}_{n} \equiv \{b_{1:n},....,b_{n-1:n-1}\}$ is 
\[
\int f^{\Xst}(\xst)p(n|N,\xst)g(\boldsymbol{b}_{n}|n,\xst)d\xst , 
\]
where $p(n|N,\xst)=C_{N,n}\left[1-F^{X|\Xst}(R|\xst)\right]^{n}\left[F^{X|\Xst}(R|\xst)\right]^{N-n}$ is the probability of observing $n$ active bidders given the number of potential bidders $N$ and UH $\xst$, and $g(\boldsymbol{b}_{n}|n,\xst)$ represents the joint PDF of the bid vector including all active bids.\footnote{Note that the identification requires the number of active bidders to be at least four. We pool bids from all auctions, including those with fewer than four active bidders, to improve estimation efficiency but rely on the auctions with $n\ge 4$ for identification.} If $n=0$, $g( \boldsymbol{0} |n,\xst)=1$ because there is no bid. If $n=1$, $g(R|1,\xst)=1$ because the bid will be $R$, as there is no reason to bid higher than the reserve price when there is only one bidder. If $2\leq n\leq N$, the joint PDF simply becomes  
\begin{eqnarray}
g(\boldsymbol{b}_{n}|n,\xst)=n!\left[1-\widetilde{F}^{X|\Xst}(b_{n-1:n})\right]\Pi_{j=1}^{n-1}\widetilde{f}^{X|\Xst}(b_{j:n}).
\end{eqnarray}


To estimate the model, we ignore the fact that we cannot identify $F^{X|\Xst}(\cdot|\tau)$ below the reserve price\footnote{Fortunately, this abstraction is barely binding for calculating the optimal reserve prices. In fact, \citet{haile2003inference} shows that as long as the existing reserve price is below the optimal, we obtain the same optimal $p^*$ by replacing $F_0$ and $f_0$ with the truncated version $F$ and $f$, respectively.} and approximate the joint density function using Berstein polynomials, $f(x,\xst; \theta) \approx \sum_{i,j}\theta_{ij}\beta_{i}(x)\beta_{j}(\xst)$, and solve the following optimization problem:\footnote{We approximate the integration by Monte Carlo simulations
\[
\int\beta_{j}(\tau)p(n_{\ell}|N_{\ell},\tau)g(\boldsymbol{b}_{\ell}|n_{\ell},\tau)d\tau\approx\frac{1}{S_{j}}\sum_{i=1}^{S_{j}}p(n_{\ell}|N_{\ell},\tau_{ij})g(\boldsymbol{b}_{\ell}|n_{\ell},\tau_{ij}),
\]
where $\tau_{ij}$ represent i.i.d. random draws from the beta density function $\beta_{j}(\cdot)$. By fixing the random draws, we make the maximization smooth in the sieve parameters $\theta$.}
\begin{eqnarray}
\max_{\theta} \quad \sum_{\ell=1}^{L}\log\left[\sum_{j}\left(\sum_{i}\theta_{ij}\right)\left\{\int\beta_{j}(\xst)p(n_{\ell}|N_{\ell},\xst;\theta)g(\boldsymbol{b}_{\ell}|n_{\ell},\xst;\theta)d\xst\right\}\right]. 
\end{eqnarray}

\subsection{Empirical Findings}

\begin{figure}[ht]
  \caption{Estimated Joint Density of UH and Bidder Value} \label{results}
  \centering
    \includegraphics[width=0.5\textwidth]{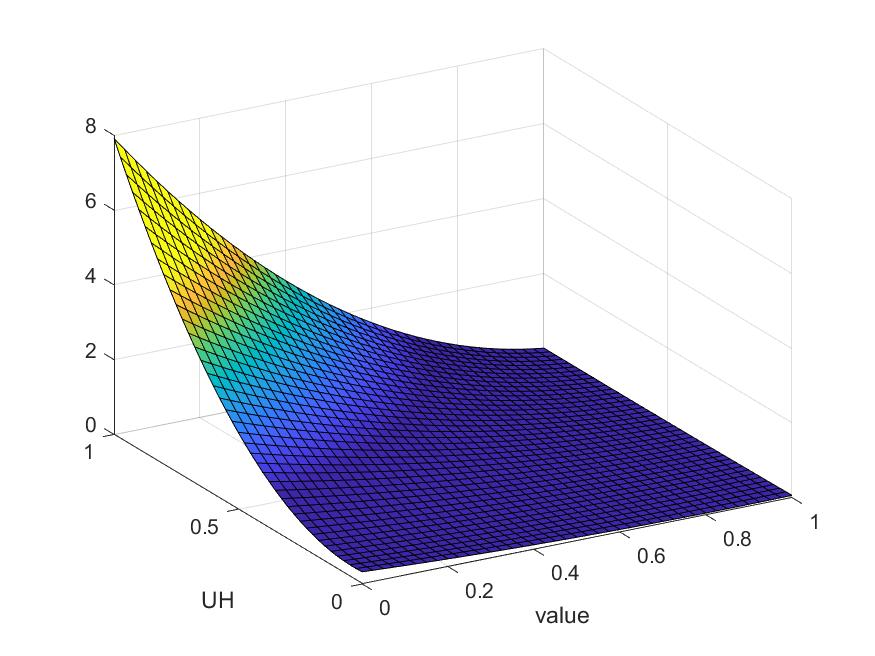}
\end{figure}

We let the number of sieve bases $J=3$. Figure \ref{results} shows the estimated joint density function of bidder value $X$ and UH $\Xst$ in homogenized and rescaled terms. Two important features are worth noting. First, the conditional densities are skewed to the left. This suggests an abundance of low willingness-to-pay amongst the potential bidders in the market, consistent with the observation that the number of registered bidders exceeds the number of actual bidders. Second, UH has important effects on bidder value. The higher $\Xst$ is, the more skewed (to the left) the density becomes. 

To demonstrate the practical use of our estimation results, we use the distribution estimated allowing for UH to calculate the optimal reserve price for each UH. Given the number of potential bidders, the optimal reserve price maximizes
\begin{eqnarray}
\pi(r,N) \!=\! N\left[1 \!-\!F(r)\right]F(r)^{N\!-\!1}(r\!-\!v_{0})+N(N\!-\!1)\int_{r}^{\overline{v}}(v\!-\!v_{0})f(v)\left[1\!-\!F(v)\right]F(v)^{N\!-\!2}dv , 
\end{eqnarray}
where $v_0$ is the seller's reserve value for keeping the item. The first term represents the seller's expected gain due to selling at the reserve price when only one value is higher than $r$, and the second term represents the gain due to selling at the second highest value when two values are higher than $r$. Its FOC leads to the following optimal reserve price 
\begin{eqnarray} \label{optreserve}
r^* = v_0 + \frac{1-F^{X|\Xst}(r^*|\xst)}{f^{X|\Xst}(r^*|\xst)},
\end{eqnarray}
which is strictly increasing in the reserve value. 
We can infer the auctioneer's reserve value from the series of judicial rules for judicial auctions issued by the supreme court. Specifically, one important rule says that the reserve price cannot be lower than $50\%$ of the appraisal value. This seems a reasonable proxy for $v_0$, i.e., $v_0 = 0.5$. 



\begin{figure}[ht]
  \caption{UH-Specific Optimal Reserve Prices} \label{optimal}
  \centering
    \includegraphics[width=0.5\textwidth]{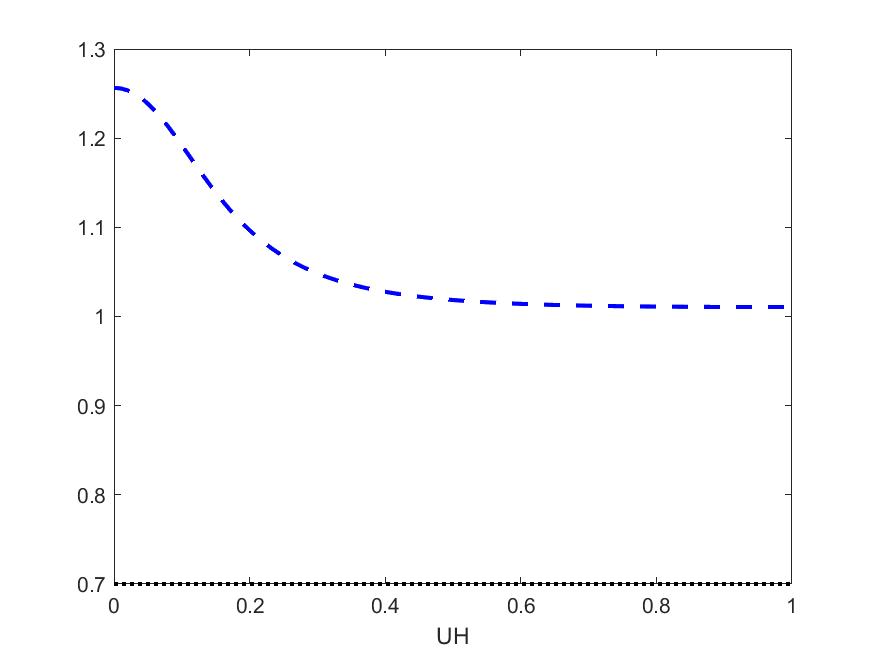}
\end{figure}
\begin{figure}[ht]
  \caption{Simple v.s. Optimal Reserve Price Schemes} \label{profit}
  \centering
    \includegraphics[width=0.5\textwidth]{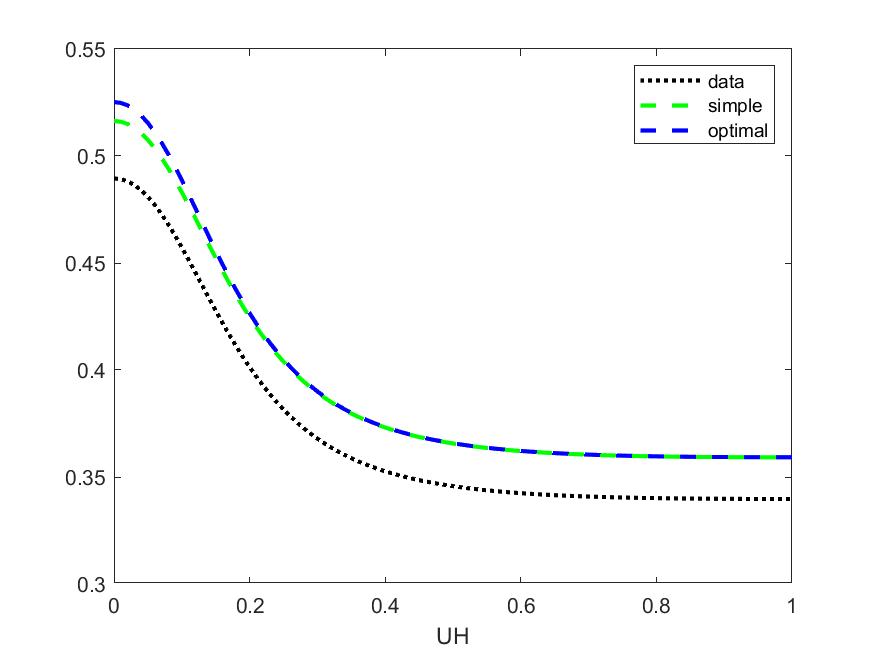}
\end{figure}

Figure \ref{optimal} shows the optimal reserve price for different levels of UH. The reserve price is strictly monotone in UH, which is consistent with the monotonicity assumption \ref{ass_order} and the estimated joint density in Figure \ref{results}. 
It is also reassuring that the optimal reserve prices are well above the current reserve price, which means that underidentification below the reserve does not prevent us from calculating the optimal reserve price.\footnote{The optimal reserve prices are still above $0.7$ with more conservative values as low as $v_0=0.1$.} In Figure \ref{profit}, the blue dashed line shows the optimal expected seller gain as a function of UH. The unconditional optimal gain $\sum_{N}p_{N}\pi(r^*, N)$ is $36.61\%$ of the appraisal value, which is $5.81\%$ higher than the current one ($34.60\%$ of the appraisal value). 

Of course, it is difficult to imagine that the seller adopts such a complex strategy. To achieve the optimal gain, the seller would need to know the UH and recover the conditional density of bidder values. Simpler strategies that require less knowledge of the value distributions are often preferable.\footnote{\citet{coey2020scalable} makes a similar point. They provide an approach to calculate optimal reserve prices without fully recovering value distributions.} We observe that the optimal reserve price is almost constant and close to one when UH is above $0.4$. Moreover, the density of UH is heavily skewed to the right (near 1). Therefore, a simple alternative to a complex UH-specific reserve price is to use the appraisal value as the reserve price. We calculate the expected revenue in this simple scheme. In this case, the unconditional expected gain is $36.59\%$ of the appraisal value, which achieves $98.85\%$ of the potential gains from the optimal reserve prices.\footnote{The appraisal value as the reserve price is nearly optimal; this finding is robust to ``large'' auctions, different seller reserve values, and alternative tuning parameters.}  

\section{Conclusion} \label{sec:conclusion}

Auction data often contain incomplete bids and miss some payoff-relevant covariates. The conventional measurement error approaches to UH are inapplicable. In this paper, we extend the analysis of \cite{hu2008instrumental} to auctions with continuous UH while accounting for incomplete bid data. Specifically, we provide point identification results for auctions with nonseparable continuous UH using consecutive order statistics of bids. We then propose sieve maximum likelihood estimators jointly for the value distribution conditional on  UH and its marginal distribution. We illustrate our methodology using a novel dataset from judicial auctions conducted by a municipal court in China. After recovering the model primitives, we propose a simple scheme that achieves nearly optimal revenue by using the appraisal value as the reserve price.


\bibliographystyle{ecta}
\bibliography{refOS}

\begin{thebibliography}{34}
\newcommand{\enquote}[1]{``#1''}
\expandafter\ifx\csname natexlab\endcsname\relax\def\natexlab#1{#1}\fi

\bibitem[\protect\citeauthoryear{Allen, Clark, Hickman, and Richert}{Allen
  et~al.}{2019}]{allen2019resolving}
\textsc{Allen, J., R.~Clark, B.~Hickman, and E.~Richert} (2019):
  \enquote{Resolving filed banks: Uncertainty, multiple bidding \& auction
  design,} Working Paper.

\bibitem[\protect\citeauthoryear{Andreyanov and Caoui}{Andreyanov and
  Caoui}{2022}]{andreyanov2020secret}
\textsc{Andreyanov, P. and E.~H. Caoui} (2022): \enquote{Secret reserve prices
  by uninformed sellers,} \emph{Quantitative Economics}, 13, 1203--1256.

\bibitem[\protect\citeauthoryear{Aradillas-L{\'o}pez, Gandhi, and
  Quint}{Aradillas-L{\'o}pez et~al.}{2013}]{aradillas2013identification}
\textsc{Aradillas-L{\'o}pez, A., A.~Gandhi, and D.~Quint} (2013):
  \enquote{Identification and inference in ascending auctions with correlated
  private values,} \emph{Econometrica}, 81, 489--534.

\bibitem[\protect\citeauthoryear{Athey and Haile}{Athey and
  Haile}{2002}]{AH2002}
\textsc{Athey, S. and P.~A. Haile} (2002): \enquote{Identification of standard
  auction models,} \emph{Econometrica}, 70, 2107--2140.

\bibitem[\protect\citeauthoryear{Chen}{Chen}{2007}]{chen2007large}
\textsc{Chen, X.} (2007): \enquote{Large sample sieve estimation of
  semi-nonparametric models,} \emph{Handbook of Econometrics}, 6, 5549--5632.

\bibitem[\protect\citeauthoryear{Chen, Fan, and Tsyrennikov}{Chen
  et~al.}{2006}]{chen2006efficient}
\textsc{Chen, X., Y.~Fan, and V.~Tsyrennikov} (2006): \enquote{Efficient
  estimation of semiparametric multivariate copula models,} \emph{Journal of
  the American Statistical Association}, 101, 1228--1240.

\bibitem[\protect\citeauthoryear{Chen and Shen}{Chen and
  Shen}{1998}]{chen1998sieve}
\textsc{Chen, X. and X.~Shen} (1998): \enquote{Sieve extremum estimates for
  weakly dependent data,} \emph{Econometrica}, 289--314.

\bibitem[\protect\citeauthoryear{Cho, Luo, and Xiao}{Cho
  et~al.}{2022}]{CLX2022}
\textsc{Cho, J., Y.~Luo, and R.~Xiao} (2022): \enquote{Deconvolution From Two
  Order Statistics,} Tech. rep., Working Paper,
  \url{https://ssrn.com/abstract=3733211}.

\bibitem[\protect\citeauthoryear{Chung}{Chung}{2000}]{chung2001course}
\textsc{Chung, K.~L.} (2000): \emph{A Course in Probability Theory, 3rd
  edition}, Academic Press.

\bibitem[\protect\citeauthoryear{Coey, Larsen, Sweeney, and Waisman}{Coey
  et~al.}{2021}]{coey2020scalable}
\textsc{Coey, D., B.~J. Larsen, K.~Sweeney, and C.~Waisman} (2021):
  \enquote{Scalable optimal online auctions,} \emph{Marketing Science}.

\bibitem[\protect\citeauthoryear{David and Nagaraja}{David and
  Nagaraja}{2004}]{david2004order}
\textsc{David, H.~A. and H.~N. Nagaraja} (2004): \emph{Order Statistics}, John
  Wiley \& Sons.

\bibitem[\protect\citeauthoryear{Freyberger and Larsen}{Freyberger and
  Larsen}{2022}]{FL2017}
\textsc{Freyberger, J. and B.~J. Larsen} (2022): \enquote{Identification in
  ascending auctions, with an application to digital rights management,}
  \emph{Quantitative Economics}, 13, 505--543.

\bibitem[\protect\citeauthoryear{Grenander}{Grenander}{1981}]{grenander1981abstract}
\textsc{Grenander, U.} (1981): \enquote{Abstract inference,} Tech. rep.

\bibitem[\protect\citeauthoryear{Guerre and Luo}{Guerre and Luo}{2022}]{GL2018}
\textsc{Guerre, E. and Y.~Luo} (2022): \enquote{Nonparametric identification of
  first-price auction with unobserved competition: a density discontinuity
  framework,} Working Paper.

\bibitem[\protect\citeauthoryear{Guerre, Perrigne, and Vuong}{Guerre
  et~al.}{2000}]{GPV2000}
\textsc{Guerre, E., I.~Perrigne, and Q.~Vuong} (2000): \enquote{Optimal
  nonparametric estimation of first-price auctions,} \emph{Econometrica}, 68,
  525--574.

\bibitem[\protect\citeauthoryear{Gut}{Gut}{2005}]{gut2005probability}
\textsc{Gut, A.} (2005): \emph{Probability: A Graduate Course}, Springer.

\bibitem[\protect\citeauthoryear{Haile and Tamer}{Haile and
  Tamer}{2003}]{haile2003inference}
\textsc{Haile, P.~A. and E.~Tamer} (2003): \enquote{Inference with an
  incomplete model of English auctions,} \emph{Journal of Political Economy},
  111, 1--51.

\bibitem[\protect\citeauthoryear{Hern{\'a}ndez, Quint, and
  Turansick}{Hern{\'a}ndez et~al.}{2020}]{HQT2019}
\textsc{Hern{\'a}ndez, C., D.~Quint, and C.~Turansick} (2020):
  \enquote{Estimation in English auctions with unobserved heterogeneity,}
  \emph{The RAND Journal of Economics}, 51, 868--904.

\bibitem[\protect\citeauthoryear{Horta{\c{c}}su and Perrigne}{Horta{\c{c}}su
  and Perrigne}{2021}]{hortaccsu2021empirical}
\textsc{Horta{\c{c}}su, A. and I.~Perrigne} (2021): \enquote{Empirical
  perspectives on auctions,} in \emph{Handbook of Industrial Organization},
  Elsevier, vol.~5, 81--175.

\bibitem[\protect\citeauthoryear{Hu}{Hu}{2008}]{H2008}
\textsc{Hu, Y.} (2008): \enquote{Identification and estimation of nonlinear
  models with misclassification error using instrumental variables: A general
  solution,} \emph{Journal of Econometrics}, 144, 27--61.

\bibitem[\protect\citeauthoryear{Hu, McAdams, and Shum}{Hu
  et~al.}{2013}]{HMS2013}
\textsc{Hu, Y., D.~McAdams, and M.~Shum} (2013): \enquote{Identification of
  first-price auctions with non-separable unobserved heterogeneity,}
  \emph{Journal of Econometrics}, 174, 186--193.

\bibitem[\protect\citeauthoryear{Hu and Schennach}{Hu and
  Schennach}{2008}]{hu2008instrumental}
\textsc{Hu, Y. and S.~M. Schennach} (2008): \enquote{Instrumental variable
  treatment of nonclassical measurement error models,} \emph{Econometrica}, 76,
  195--216.

\bibitem[\protect\citeauthoryear{Komarova}{Komarova}{2013}]{komarova2013new}
\textsc{Komarova, T.} (2013): \enquote{A new approach to identifying
  generalized competing risks models with application to second-price
  auctions,} \emph{Quantitative Economics}, 4, 269--328.

\bibitem[\protect\citeauthoryear{Krasnokutskaya}{Krasnokutskaya}{2011}]{K2011}
\textsc{Krasnokutskaya, E.} (2011): \enquote{Identification and estimation of
  auction models with unobserved heterogeneity,} \emph{The Review of Economic
  Studies}, 78, 293--327.

\bibitem[\protect\citeauthoryear{Li, Perrigne, and Vuong}{Li
  et~al.}{2000}]{li2000conditionally}
\textsc{Li, T., I.~Perrigne, and Q.~Vuong} (2000): \enquote{Conditionally
  independent private information in OCS wildcat auctions,} \emph{Journal of
  Econometrics}, 98, 129--161.

\bibitem[\protect\citeauthoryear{Li and Vuong}{Li and Vuong}{1998}]{LV1998}
\textsc{Li, T. and Q.~Vuong} (1998): \enquote{Nonparametric estimation of the
  measurement error model using multiple indicators,} \emph{Journal of
  Multivariate Analysis}, 65, 139--165.

\bibitem[\protect\citeauthoryear{Lu and Perrigne}{Lu and
  Perrigne}{2008}]{lu2008estimating}
\textsc{Lu, J. and I.~Perrigne} (2008): \enquote{Estimating risk aversion from
  ascending and sealed-bid auctions: The case of timber auction data,}
  \emph{Journal of Applied Econometrics}, 23, 871--896.

\bibitem[\protect\citeauthoryear{Luo}{Luo}{2020}]{Luo2018}
\textsc{Luo, Y.} (2020): \enquote{Unobserved heterogeneity in auctions under
  restricted stochastic dominance,} \emph{Journal of Econometrics}, 216,
  354--374.

\bibitem[\protect\citeauthoryear{Luo and Xiao}{Luo and
  Xiao}{2022}]{luo2020identification}
\textsc{Luo, Y. and R.~Xiao} (2022): \enquote{Identification of Auction Models
  Using Order Statistics,} Working Paper,
  \url{https://arxiv.org/abs/2205.12917}.

\bibitem[\protect\citeauthoryear{Mbakop}{Mbakop}{2017}]{M2017}
\textsc{Mbakop, E.} (2017): \enquote{Identification of auctions with incomplete
  bid data in the presence of unobserved heterogeneity,} Tech. rep., Working
  Paper.

\bibitem[\protect\citeauthoryear{Menzel and Morganti}{Menzel and
  Morganti}{2013}]{menzel}
\textsc{Menzel, K. and P.~Morganti} (2013): \enquote{Large sample properties
  for estimators based on the order statistics approach in auctions,}
  \emph{Quantitative Economics}, 4, 329--375.

\bibitem[\protect\citeauthoryear{Shen}{Shen}{1997}]{shen1997methods}
\textsc{Shen, X.} (1997): \enquote{On methods of sieves and penalization,}
  \emph{The Annals of Statistics}, 2555--2591.

\bibitem[\protect\citeauthoryear{Van~der Vaart and Wellner}{Van~der Vaart and
  Wellner}{1996}]{van199}
\textsc{Van~der Vaart, A.~W. and J.~A. Wellner} (1996): \emph{Weak Convergence
  and Empirical Processes with Application to Statistics}, New York, Springer.

\bibitem[\protect\citeauthoryear{Wu and Zhang}{Wu and
  Zhang}{2012}]{wu2012partially}
\textsc{Wu, Y. and Y.~Zhang} (2012): \enquote{Partially monotone tensor spline
  estimation of the joint distribution function with bivariate current status
  data,} \emph{The Annals of Statistics}, 40, 1609--1636.

\end{thebibliography}


\begin{thebibliography}{11}
\newcommand{\enquote}[1]{``#1''}
\expandafter\ifx\csname natexlab\endcsname\relax\def\natexlab#1{#1}\fi

\bibitem[\protect\citeauthoryear{Compiani, Haile, and Sant’Anna}{Compiani
  et~al.}{2020}]{compiani2018common}
\textsc{Compiani, G., P.~Haile, and M.~Sant’Anna} (2020): \enquote{Common
  values, unobserved heterogeneity, and endogenous entry in US offshore oil
  lease auctions,} \emph{Journal of Political Economy}, 128, 3872--3912.

\bibitem[\protect\citeauthoryear{David and Nagaraja}{David and
  Nagaraja}{2004}]{david2004order}
\textsc{David, H.~A. and H.~N. Nagaraja} (2004): \emph{Order Statistics}, John
  Wiley \& Sons.

\bibitem[\protect\citeauthoryear{Dunford and Schwartz}{Dunford and
  Schwartz}{1971}]{DS1971}
\textsc{Dunford, N. and J.~Schwartz} (1971): \enquote{Linear operators,}
  \emph{New York: Wiley}.

\bibitem[\protect\citeauthoryear{Ghosal}{Ghosal}{2001}]{ghosal2001}
\textsc{Ghosal, S.} (2001): \enquote{Convergence rates for density estimation
  with {Bernstein} polynomials,} \emph{The Annals of Statistics}, 29,
  1264--1280.

\bibitem[\protect\citeauthoryear{Gut}{Gut}{2005}]{gut2005probability}
\textsc{Gut, A.} (2005): \emph{Probability: A Graduate Course}, Springer.

\bibitem[\protect\citeauthoryear{Hu and Schennach}{Hu and
  Schennach}{2008}]{hu2008instrumental}
\textsc{Hu, Y. and S.~M. Schennach} (2008): \enquote{Instrumental variable
  treatment of nonclassical measurement error models,} \emph{Econometrica}, 76,
  195--216.

\bibitem[\protect\citeauthoryear{Kong}{Kong}{2020}]{kong2020not}
\textsc{Kong, Y.} (2020): \enquote{Not knowing the competition: evidence and
  implications for auction design,} \emph{The RAND Journal of Economics}.

\bibitem[\protect\citeauthoryear{Petrone and Wasserman}{Petrone and
  Wasserman}{2002}]{petrone}
\textsc{Petrone, S. and L.~Wasserman} (2002): \enquote{Consistency of Bernstein
  polynomial posteriors,} \emph{Journal of the Royal Statistical Society:
  Series B (Statistical Methodology)}, 64, 79--100.

\bibitem[\protect\citeauthoryear{Schumacker}{Schumacker}{1981}]{sch1981}
\textsc{Schumacker, L.} (1981): \emph{Spline Functions: Basic Theory}, New
  York: Wiley Interscience.

\bibitem[\protect\citeauthoryear{Van~der Vaart}{Van~der Vaart}{1998}]{van1998}
\textsc{Van~der Vaart, A.~W.} (1998): \emph{Asymptotic statistics}, Cambridge,
  Cambridge University.

\bibitem[\protect\citeauthoryear{Zeng}{Zeng}{2005}]{zeng2005}
\textsc{Zeng, D.} (2005): \enquote{Likelihood approach for marginal
  proportional hazards regression in the presence of dependent censoring,}
  \emph{The Annals of Statistics}, 33, 501--521.

\end{thebibliography}

\end{document}




	\section*{Online Supplement (not for publication)}
	
	This Online Supplement contains three parts. First, we provide the proof details omitted in the identification section. Second, we show the consistency of the sieve MLEs and further investigate the consistency of B-splines and Bernstein polynomials. Moreover, we conduct Monte Carlo simulations to demonstrate the finite-sample property of Bernstein polynomials. Finally, we extend our identification results to (any) four order statistics. 
	
	\setcounter{page}{1}
	\setcounter{subsection}{0}
	
	\setcounter{equation}{0}
	\setcounter{ass}{0}
	\setcounter{theorem}{0}
	\setcounter{lemma}{0}
	\setcounter{corollary}{0}
	
	\renewcommand{\theequation}{S\arabic{equation}}
	\renewcommand{\thefigure}{S\arabic{figure}}
	\renewcommand{\thelemma}{S\arabic{lemma}}
	\renewcommand{\thetheorem}{S\arabic{theorem}}
	\renewcommand{\theass}{S\arabic{ass}}
	\renewcommand{\thecorollary}{S\arabic{corollary}}
	
	\renewcommand{\bibnumfmt}[1]{[S#1]}
	\renewcommand{\citenumfont}[1]{S#1}
	
	\setcounter{section}{0}
	\renewcommand{\thesection}{S.\arabic{section}}

	\section{Identification Proofs} \label{sec:Idproof}
	
\subsection{Derivation of Equation (2)}

Ignoring UH for the moment, the joint distribution of any three order statistics does not have a multiplicatively separable structure, i.e.,  $f_{r,s,t:n}(x,y,z)  \sim f(x)f(y)f(z) [F(x)]^{r-1} [F(y)-F(x)]^{s-r-1}[F(z)-F(y)]^{t-s-1}[1-F(z)]^{n-t}$, where $r<s<t$; see \cite{david2004order}. Considering consecutive order statistics, we have 
    \begin{eqnarray}
f_{r-2, r-1,r:n}(x,y,z)& = & \frac{n!}{(r-3)!(n-r)!} [F(x)]^{r-3} f(x) f(y) [1-F(z)]^{n-r}f(z) \mathbbm{1}(x\le y\le z )\notag \\
    &\equiv& c_{r,n} f_{r-2:r-2}(x) \cdot f(y)  \cdot f_{1:n-r+1}(z)  \cdot \mathbbm{1}(x\le y\le z ), 
    \end{eqnarray}
where $c_{r,n}\equiv  \frac{n!}{(r-2)!(n-r+1)!}$, $f_{r-2:r-2}(x)$ is the PDF of the top-order statistic of a sample of size $r-2$, and $f_{1:n-r+1}(z)$ is the PDF of the bottom-order statistic of a sample of size $n-r+1$, and $\mathbbm{1}(\cdot)$ is an indicator function. Bringing back UH gives Equation (2) by the law of total probability. 
	
\subsection{Derivation of the Equivalence of Operators (Equation (4))} 
	
	We derive the equivalence of the operators as follows. Specifically, for any given $x \in \mA_l$ and $y \in \mA_m$, we have 
	\begin{eqnarray}
		[J_{y}g](x) &\equiv & \int_{\mathcal{X}_h} f_{r-2,r-1,r:n}(x, y, z) g(z)dz \notag \\
		&=& \int_{\mathcal{X}_h} c_{r,n} \cdot  \int_{\XXst}  f^{X|\Xst}_{r-2:r-2}(x|\xst) f^{X|\Xst}(y|\xst) f^{X|\Xst}_{1:n-r+1}(z|\xst)  f^{\Xst}(\xst) d \xst g(z) dz \notag  \\
		&=&  \int_{\XXst}  f^{X|\Xst}_{r-2:r-2}(x|\xst)  c_{r,n} f^{X|\Xst}(y|\xst)  f^{\Xst}(\xst) \left\{\int_{\mathcal{X}_h} f^{X|\Xst}_{1:n-r+1}(z|\xst)g(z) dz\right\} d \xst \notag \\
		&=& \int_{\XXst}  f^{X|\Xst}_{r-2:r-2}(x|\xst)  c_{r,n} f^{X|\Xst}(y|\xst)  f^{\Xst}(\xst) [H_{X_{1:n-r+1}|\Xst}g] (\xst) d \xst \notag \\
		&\equiv & \int_{\XXst}  f^{X|\Xst}_{r-2:r-2}(x|\xst)  [\Delta_{X=y,\Xst}H_{X_{1:n-r+1}|\Xst}g] (\xst) d \xst \notag \\
		&=& [L_{X_{r-2:r-2}|\Xst} \Delta_{X=y,\Xst}H_{X_{1:n-r+1}|\Xst}g](x),
		\label{eq_joint2}
	\end{eqnarray}
	which implies that the operators from both sides are equivalent. 

	
	\subsection{Injectivity Condition}

Here, we prove that (i) the family of distributions $\{f^{X | \Xst}_{r - 2: r - 2}(x | \tau): x \in \mathcal{X}_l \}$ is complete when the following condition is satisfied: if
$$
\int_{\XXst} g(\xst) F^{X | \Xst }( x | \xst)^{r-2} d\xst = 0
$$
holds for all $x \in \mathcal X_l$, 
we have $g(\xst)=0$, $\forall \xst \in \XXst$; (ii)  if the original distribution family is complete, i.e.,$\{f^{X|\Xst}_{1:n-r+1}(x| \xst):\xst \in \XXst \}$ over $\mathcal{L}^2(\mathcal{X})$ is complete, there exists at least one division of the support such that operator $H_{X_{1:n-r+1}|\Xst}$ is injective.
	\begin{proof} 
	(i) The proof is achieved by contraction. Specifically, if the family distribution
$\{f_{r-2:r-2}^{X | \Xst}(x | \tau): x \in \mathcal{X}_l \}$ is not complete over $\mathcal{L}^2(\XXst)$, there exists $\tilde g(\xst) \neq 0$, such that $$
\int_{\XXst} (r-2) \tilde  g(\xst)f^{X | \Xst }( x | \xst) F^{X | \Xst }( x | \xst)^{r-3} d\xst = 0.
$$
We can then integrate the component on the left-hand side of the above equation with respect to the support of $x$, leading to
$$
\int_{\XXst} \int_x (r-2) \tilde  g(\xst)f^{X | \Xst }( x | \xst) F^{X | \Xst }( x | \xst)^{r-3}dx d\xst = \int_{\XXst} \tilde g(\xst) F^{X | \Xst }( x | \xst)^{r-2} d\xst = 0,
$$
which contradicts the provided condition. Therefore, we have proved that the family of distributions $\{f^{X | \Xst}_{r - 2: r - 2}(x | \tau): x \in \mathcal{X}_l \}$ is complete given the condition in ($i$) holds.   

\bigskip

(ii) We only need to show that there exists at least a division, $\{\mathcal X_l, \mathcal X_m, \mathcal X_h\}$, 
such that the unique solution to $\int_{\mathcal X_h} g(x) f^{X | \Xst }_{1: n-r +1}( x | \xst) dx = 0$ for all $\tau \in \mathcal T$ is $g = 0$. 
We show this by contradiction. Suppose, for any division $\{\mathcal X_l, \mathcal X_m, \mathcal X_h\}$, there must exist a nonzero $g$ such that
$$
\int_{\mathcal X_h} g(x) f^{X | \Xst }_{1: n-r +1}( x | \xst) dx = 0,
$$
for any $\tau \in \mathcal T$. 
We construct a new function $\tilde g$, which equals to $g$ on $\mathcal X_h$ and 0 otherwise. Obviously, $\tilde g  \neq  0$, but we have 
$$
\int_{\mathcal X} \tilde g(x) f^{X | \Xst }_{1: n-r +1}(x | \xst) dx = \int_{\mathcal X_h} g(x) f^{X | \Xst }_{1: n-r +1}( x | \xst) dx= 0, 
$$
which holds for every $\xst \in \XXst$. This is contradictory to the assumption that the family of distribution $\{f^{X | \Xst}_{1:n- r +1}( x |\xst): \xst \in \XXst\}$ is complete on $\mathcal L^2(\mathcal X)$. 

\end{proof} 
	
	\subsection{Proof of Lemma \ref{le-bijective}}
	
	We first show that 
	$L_{X_{r-2:r-2}|\Xst}^{\ast}$, the adjoint operator of $L_{X_{r-2:r-2}|\Xst}$, is injective under Assumptions 1-3. To prove this,  for any $g_1 \in \mL^2(\XXst)$ and $g_2 \in \mL^2(\mA_l)$, we have, from the definition of the adjoint operator,
	\begin{equation}
		\langle L_{X_{r-2:r-2}|\Xst} g_1, g_2 \rangle_{\mL^2(\mA_l)} = \langle  g_1, L_{X_{r-2:r-2}|\Xst}^{\ast}g_2 \rangle_{\mL^2(\XXst)},
		\label{eq-adjoint}
	\end{equation}
	where $\langle \cdot, \cdot \rangle$ is the inner product defined on the $\mL^2$ space. The left-hand side of the equation above can be further expressed as 
	$$
	\int_{\mA_l} \int_{\XXst} f^{X|\Xst}_{r-2:r-2}(x|\xst) g_1(\xst)d\xst  g_2(x)   dx.
	$$
	Obviously, Equation \eqref{eq-adjoint} holds if and only if
	$$
	[L_{X_{r-2:r-2}|\Xst}^{\ast}g_2](\xst) = \int_{\mA_l}  f^{X|\Xst}_{r-2:r-2}(x|\xst) g_2(x) dx. 
	$$
	The right-hand side of the equation above can be rewritten as
	$$
	\int_{\mA_l}  f^{\Xst \mid X_{r-2: r-2}}(\xst| x) \cdot \frac{f^{X_{r-2:r-2}}(x)}{f^{\Xst}(\xst)} g_2(x) dx. 
	$$
	Thus, the adjoint operator $L_{X_{r-2:r-2}|\Xst}^{\ast}$ is injective, according to the equivalent of a family of distributions and given that $L_{\Xst|X_{r-2:r-2}}$ is injective.  
	
	Given that $L_{X_{r-2:r-2}|\Xst}$ is an operator from one Hilbert space to another Hilbert space,  its null space is the complement of the closure of
	the range of $L_{X_{r-2:r-2}|\Xst}^{\ast}$, denoted as $\overline{\mR(L_{X_{r-2:r-2}|\Xst}^{\ast})}$. Therefore, $L_{X_{r-2:r-2}|\Xst}$ is injective when it is viewed as a mapping of $\overline{\mR(L_{X_{r-2:r-2}|\Xst}^{\ast})}$ to $\mL^2(\mA_l)$. It follows that $L_{X_{r-2:r-2}|\Xst}^{-1}$ exists.
	
	Moreover,  the closure of
	the range of $L_{X_{r-2:r-2}|\Xst}$, $\overline{\mR(L_{X_{r-2:r-2}|\Xst})}$, is the orthogonal complement of the null space of $L_{X_{r-2:r-2}|\Xst}^{\ast}$. This null space is $\{0\}$, because $L_{X_{r-2:r-2}|\Xst}^{\ast}$ is injective. Consequently, $\overline{\mR(L_{X_{r-2:r-2}|\Xst})} = \mL^2(\mA_l)$. Therefore,  $L_{X_{r-2:r-2}|\Xst}^{-1}$ is  defined over a dense subset of $\mL^2(\mA_l)$. 
	\QEDA
	
	\subsection{Derivation of the Main Identification Equation (Equation (5))} 
	
	We first derive the main equation for identification in the following. Specifically, we have the following equations for any two values of $y$:
	\begin{eqnarray}
		J_{y_1} &=& L_{X_{r-2:r-2}|\Xst} \Delta_{X=y_1,\Xst}H_{X_{1:n-r+1}|\Xst}, \label{eq_joint4}\\
		J_{y_2} &=& L_{X_{r-2:r-2}|\Xst} \Delta_{X=y_2,\Xst}H_{X_{1:n-r+1}|\Xst}\label{eq_joint5}.
	\end{eqnarray}
	From Equation \eqref{eq_joint5}, we obtain the following equivalence of the operator 
	\begin{eqnarray}
		\Delta^{-1}_{X=y_2,\Xst}L^{-1}_{X_{r-2:r-2}|\Xst} J_{y_2} &=&H_{X_{1:n-r+1}|\Xst},\label{eq_joint6}
	\end{eqnarray}
	which holds for the same domain $\mathcal{G} (\mathcal{X}_h)$  as in Equation \eqref{eq_joint5} because the inverse operators ($L^{-1}_{X_{r-2:r-2}|\Xst}$ and $\Delta^{-1}_{X=y_2,\Xst}$) were applied from the left side of Equation \eqref{eq_joint5} in the correct order. 
	
	We plug this equation back into Equation \eqref{eq_joint4}, leading to the following equation:
	\begin{eqnarray}
		J_{y_1} &=& L_{X_{r-2:r-2}|\Xst} \Delta_{X=y_1,\Xst} \Delta^{-1}_{X=y_2,\Xst}L^{-1}_{X_{r-2:r-2}|\Xst} J_{y_2}. \label{eq_joint7}
	\end{eqnarray}
	Note that the operator $J_{y_2}$ is injective, guaranteed by the injection of operators \\ $L_{X_{r-2:r-2}|\Xst}$ and $H_{X_{1:n-r+1}|\Xst}$. Thus, we obtain the main equation for identification by right multiplying the inverse of the operator $J_{y_2}$:
	$$
	J_{y_1} J^{-1}_{y_2} = L_{X_{r-2:r-2}|\Xst} \Delta_{X=y_1,\Xst} \Delta^{-1}_{X=y_2,\Xst}L^{-1}_{X_{r-2:r-2}|\Xst}.
	$$
	
	\subsection{Proof of Theorem 1} 
	
	This proof consists of the following steps: 1) we identify the conditional marginal distribution $f^{X|\Xst}(x|\xst)$ in segment $\mathcal{X}_l$ up to scales, ordering, and location, 2) we identifythe conditional marginal  distribution $f^{X|\Xst}(x|\xst)$ in segment $x \in \mA_h$ up to scales, ordering, and location, 3) we identify the conditional marginal distribution $f^{X|\Xst}(x|\xst)$ in segment $x \in \mA_m$ up to scales, ordering, and location. Moreover, the ordering and location of UH are consistently matched across all three segments, but scales may vary; 4) we then proceed to pin down the scales and exact location of UH, and 5) we identify the marginal distribution $f^{\Xst}(\xst)$. 
	
	\paragraph{Step 1}
	The identification of step 1) mainly consists of the following three steps. First, we show that Equation \eqref{eq_joint8} admits a unique representation. Second, we provide sufficient conditions under which  the eigen-decomposition of the component on the left-hand side of Equation \eqref{eq_joint8} is unique. Thus, the main equation for identification generates unique eigenfunctions $f^{X|\Xst}_{r-2: r -2} (x | \Xst)$ for $x \in \mA_l$. Lastly, we show that the conditional marginal distribution $f^{X|\Xst}(x|\xst)$ is identified in the same domain using the one-to-one mapping between the distribution of an order statistic and its parent distribution in a given segment. The detailed proofs for each step are as follows. 
	
	First, we show that  Equation \eqref{eq_joint8} admits a unique representation. 
	The operator on the left-hand side of Equation \eqref{eq_joint8}, $J_{y_1} J^{-1}_{y_2} $, is determined by the densities of the observed three consecutive order statistics and so can be viewed as known. This equation implies that $J_{y_1} J^{-1}_{y_2}$ admits a spectral decomposition. More specifically, the ``diagonal elements" of operator $\Delta_{X=y_1,\Xst} \Delta^{-1}_{X=y_2,\Xst}$, i.e., $\left\{\frac{f^{X|\Xst}(y_1 | \xst)}{f^{X|\Xst}(y_2 | \xst)}\right\}$ for given $y_1, y_2$ and for all $\xst$, and the kernel of the integral operator $L_{X_{r-2:r-2}|\Xst}$, i.e., $\{f^{X|\Xst}_{r-2:r-2}(.| \xst)\}$ for all $\tau$, are the eigenvalues and  eigenfunctions of operator $J_{y_1} J^{-1}_{y_2} $, respectively. 
	
We follow the sufficient and necessary conditions provided in Theorem XV.4.5 in \cite{DS1971} for the existence of a unique representation via spectral decomposition of a linear operator. Specifically, if a bounded and linear operator $A$ can be written as $A = U + V$, and $U$ is an operator represented as
	\begin{eqnarray}
		U = \int_{\sigma} \lambda Q(d\lambda),
		\label{eq-proj}
	\end{eqnarray}
	where $Q$ is a projection-valued measure with the support being spectrum $\sigma$, a subset of the complex field, and $V$ is a ``quasi-nilpotent" operator computing with $U$, then this representation is unique. We apply this general result to our problem where $A = J_{y_1} J^{-1}_{y_2}$, $\sigma \subset \mathbb{R}$, and $V = 0$. Thus, we need to prove that linear operator $A = J_{y_1} J^{-1}_{y_2}$ is bounded, which can be accomplished by showing that the spectrum of operator $L_{X_{r-2:r-2}|\Xst} \Delta_{X=y_1,\Xst} \Delta^{-1}_{X=y_2,\Xst}L^{-1}_{X_{r-2:r-2}|\Xst}$ is a compact set, so that this operator is bounded because it is positive and its spectrum is a compact set.
	
First, we prove that the spectrum of $L_{X_{r-2:r-2}|\Xst}\mathcal{D}_{y_1, y_2 }L_{X_{r-2:r-2}|\Xst}^{-1}$ is a compact set. Because $\frac{f^{X|\Xst}(y_1 | \cdot)}{f^{X|\Xst}(y_2 | \cdot)}$ is continuous and bounded by Assumption 2, denote the range of $\left\{\frac{f^{X|\Xst}(y_1 | \xst)}{f^{X|\Xst}(y_2 | \xst)}: \xst \in \XXst\right\}$ as $\mathcal{I} = [\lambda_1(y_1, y_2), \lambda_2(y_1, y_2)]$. Denote the spectrum of $L_{X_{r-2:r-2}|\Xst}\mathcal{D}_{y_1, y_2 }L_{X_{r-2:r-2}|\Xst}^{-1}$ as $\sigma$.
We will show that $\mathcal{I} = \sigma$, which consists of the following two parts.

(i) $\sigma \subset \mathcal I$. Define $\mathcal{D}_{y_1, y_2}$ as an operator from 
	$\mL^2(\XXst)$ to $\mL^2(\XXst)$ such that 
	$$
	(\mathcal{D}_{y_1, y_2} g)(\tau) = \frac{f^{X|\Xst}(y_1 | \xst)}{f^{X|\Xst}(y_2 | \xst)} g(\tau)
	$$
	for any $g \in \mL^2(\XXst)$. Let $\sigma_{\mathcal{D}}$ denote the spectrum of $\mathcal{D}_{y_1, y_2 }$.
	By definition, 
	$$
	[( \mathcal{D}_{y_1, y_2} - \lambda \cdot id) g](\tau) = \frac{f^{X|\Xst}(y_1 | \xst)}{f^{X|\Xst}(y_2 | \xst)} g(\tau) - \lambda g(\tau) = \left\{\frac{f^{X|\Xst}(y_1 | \xst)}{f^{X|\Xst}(y_2 | \xst)} - \lambda\right\} g(\tau). $$

For any $\lambda \notin \mathcal{I}$,  $f^{X|\Xst}(y_1 | \xst) f^{X|\Xst}(y_2 | \xst)^{-1} - \lambda \neq 0$ for any $\tau \in \mathcal{T}$. Following the last displayed equation, $[( \mathcal{D}_{y_1, y_2} - \lambda \cdot id) g](\tau) \equiv 0$ implies that $g \equiv 0$. Hence $\lambda \notin \sigma_{\mathcal{D}}$. 

 
 Next, we show that $L_{X_{r-2:r-2}|\Xst}\mathcal{D}_{y_1, y_2 }L_{X_{r-2:r-2}|\Xst}^{-1} -  \lambda \cdot id$ is invertible if $\lambda \notin \mathcal{I}$.  Actually,  given that 
	$$
	L_{X_{r-2:r-2}|\Xst}\mathcal{D}_{y_1, y_2 }L_{X_{r-2:r-2}|\Xst}^{-1} -  \lambda \cdot id = 	L_{X_{r-2:r-2}|\Xst}(\mathcal{D}_{y_1, y_2 } -  \lambda \cdot id) L_{X_{r-2:r-2}|\Xst}^{-1}, 
	$$
 which is invertible because $(\mathcal{D}_{y_1, y_2 } -  \lambda \cdot id)$ is invertible from $\lambda \notin \sigma_{\mathcal{D}}$, it is thus bijective if $\lambda \notin \mathcal{I}$; namely $\lambda \notin \sigma$ if $\lambda \notin \mathcal{I}$. It follows that $\sigma \subset  \mathcal{I}$.
 
(ii) $\mathcal{I} \subset \sigma$. It is straightforward to verify that, for any $\lambda \in \mathcal{I}$, we can find a non-vanishing $g \in \mL(\mA_l)$ such that 
 $
 (L_{X_{r-2:r-2}|\Xst}\mathcal{D}_{y_1, y_2 }L_{X_{r-2:r-2}|\Xst}^{-1}- \lambda \cdot id) g \equiv 0.
 $ In fact, because $\lambda \in \mathcal{I}$, there exists $\xst_0$ such that
 \[
\frac{f^{X|\Xst}(y_1 | \xst_0)}{f^{X|\Xst}(y_2 | \xst_0)}= \lambda,  
 \]
which implies that $(\mathcal{D}_{y_1, y_2 } -  \lambda \cdot id)|_{\xst = \xst_0} =0$. Define $g(\xst) = 1$ if $\xst = \xst_0$, and $=0$ otherwise. 

 
 Combining (i) and (ii), we have $\sigma = \mathcal{I}$.

Second, we show that $L_{X_{r-2:r-2}|\Xst}\mathcal{D}_{y_1, y_2 }L_{X_{r-2:r-2}|\Xst}^{-1}$ is a bounded operator. By definition, we have
 \begin{align*}
     \|L_{X_{r-2:r-2}|\Xst}\mathcal{D}_{y_1, y_2 }L_{X_{r-2:r-2}|\Xst}^{-1}\| & = \sup_{\|u\|_{\mathcal L^2} = 1} \langle L_{X_{r-2:r-2}|\Xst}\mathcal{D}_{y_1, y_2 }L_{X_{r-2:r-2}|\Xst}^{-1} u, L_{X_{r-2:r-2}|\Xst}\mathcal{D}_{y_1, y_2 }L_{X_{r-2:r-2}|\Xst}^{-1}u\rangle_{\mathcal L^2} \\
     & = \sup_{\|u\|_{\mathcal L^2} = 1} \int_{\mathcal I} \lambda^2 d \mu_{u, u}(\lambda) \\
     & \leq \sup_{\lambda \in \mathcal{I}} |\lambda|^2 \sup_{\|u\|_{\mathcal L^2} = 1}\|\mu_{u, u}\| \leq \sup_{\lambda \in \mathcal{I}} |\lambda|^2 < \infty. 
 \end{align*}
The last inequality follows from $\mathcal{I}$ and is compact. Therefore, $L_{X_{r-2:r-2}|\Xst}\mathcal{D}_{y_1, y_2 }L_{X_{r-2:r-2}|\Xst}^{-1}$ is a bounded operator. 
 
As a result, we prove that $J_{y_1}J_{y_2}^{-1}$ is bounded because 
 \[J_{y_1}J_{y_2}^{-1} = L_{X_{r-2:r-2}|\Xst}\mathcal{D}_{y_1, y_2 }L_{X_{r-2:r-2}|\Xst}^{-1}.\]

We define the projection-valued measure $Q$ in the following way: For any 
	$\Lambda \subset \mathbb{R}$,
	$$
	Q(\Lambda) = L_{X_{r-2:r-2}|\Xst} I_{\Lambda}L_{X_{r-2:r-2}|\Xst} ^{-1},
	$$
	where operator $I_{\Lambda}$ is defined as
	$$
	[I_{\Lambda}g](\xst) = \mathbbm{1}\left(\frac{f^{X|\Xst}(y_1 | \xst)}{f^{X|\Xst}(y_2 | \xst)}\in \Lambda\right) g(\xst),
	$$
	 we want to show that $\int_{\sigma} \lambda Q(d\lambda) = L_{X_{r-2:r-2}|\Xst} \Delta_{X=y_1,\Xst} \Delta^{-1}_{X=y_2,\Xst}L^{-1}_{X_{r-2:r-2}|\Xst}$. Based on the definition of $Q$, we have
	\begin{align*}
		\int_{\sigma} \lambda Q(d\lambda) & = \int_{\sigma} \lambda \left(\frac{d}{d\lambda}Q((-\infty, \lambda])\right) d\lambda\\
		& =  L_{X_{r-2:r-2}|\Xst} \left(\int_{\sigma} \lambda \frac{d \mathbbm{1}_{(-\infty, \lambda]}}{d\lambda} d\lambda\right)L_{X_{r-2:r-2}|\Xst} ^{-1}. 
	\end{align*}
	To find the operator $\int_{\sigma} \lambda \frac{d \mathbbm{1}_{(-\infty, \lambda]}}{d\lambda} d\lambda$, we investigate its evaluation when operating on a function $g$. That is,
	\begin{align*}
		\left[\int_{\sigma} \lambda \frac{d \mathbbm{1}_{(-\infty, \lambda]}}{d\lambda} d\lambda g\right](\xst) & = \int_{\sigma} \lambda \frac{d}{d\lambda} \mathbbm{1} \left(\frac{f^{X|\Xst}(y_1 | \xst)}{f^{X|\Xst}(y_2 | \xst)}\in (-\infty, \lambda] \right) g(\xst) d\lambda \\
		& = \int_{\sigma} \lambda \delta \left(\lambda - \frac{f^{X|\Xst}(y_1 | \xst)}{f^{X|\Xst}(y_2 | \xst)} \right)
		g(\xst) d\lambda \\
		& = \frac{f^{X|\Xst}(y_1 | \xst)}{f^{X|\Xst}(y_2 | \xst)} g(\xst) = [\Delta_{X=y_1,\Xst} \Delta^{-1}_{X=y_2,\Xst}g](\xst), 
	\end{align*} 
	where we have used the Dirac delta function $\delta$ satisfying the property that
	$\int \delta(x - x_0) h(x) dx = h(x_0)$ for any function $h$ continuous at $x = x_0$. 
	It follows that $\int_{\sigma} \lambda Q(d\lambda) = L_{X_{r-2:r-2}|\Xst} \Delta_{X=y_1,\Xst} \Delta^{-1}_{X=y_2,\Xst}L^{-1}_{X_{r-2:r-2}|\Xst}$. 
	
	Second, we provide sufficient conditions under which the eigen-decomposition given in Equation \eqref{eq_joint8} is unique. Note that the uniqueness of the representation in Equation \eqref{eq-proj} does not necessarily indicate that spectral decomposition of $ J_{y_1} J^{-1}_{y_2}$ in Equation \eqref{eq_joint8} is unique. This uniqueness problem is similar in spirit to a unique eigen-decomposition of a square matrix in  the following two respects:
	
	1. There is a unique eigen-space $S_{\lambda}$ spanning eigenfunctions corresponding to each eigenvalue $\lambda$. However, there are many different ways to select a basis for this space.
	\begin{itemize}
		\item[1a. ] Each basis can be multiplied by a constant. Unlike in  \cite{hu2008instrumental}, where the scaling problem of the eigenfunction $\{f^{X|\Xst}_{r-2:r-2}(\cdot | \xst)\}$ could be addressed using the fact that total probability is 1, we cannot apply the same logic in our context because the eigenfunction in our context is density functions in a segment of the full support.
		
		\item[1b. ] If the dimension of one eigen-space is larger than 1, then a new eigen-function can be constructed through a linear combination of original basis functions. 
	\end{itemize}
	
	2. Our economic model indexes the eigenvalues by $\xst$ and establishes the one-to-one mapping between eigenvalues and eigen-space. However, other methods can be used to index eigenvalues. In other words, if we use $\lambda(\xst)$ to denote the mapping between $\xst$ and $\lambda$ (and hence $S_{\lambda(\xst)})$, the choice of $\lambda(\xst)$ is not unique. The supplementary material of \cite{hu2008instrumental} shows non-uniqueness of indexing
	eigenvalues in some scenarios. 
	%
	
	We harness Assumption \ref{ass_distinct} to address issue (1b). Note that the integral operator $L_{X_{r-2:r-2}|\Xst}$, with the kernel being the eigenfunction, depends on neither $y_1$ nor $y_2$, but the eigenvalues $\frac{f^{X|\Xst}(y_1 | \xst)}{f^{X|\Xst}(y_2 | \xst)}$  do. If there exist two different values of $y$, say $y_1$ and $y_2$, such that there are two eigenfunctions $f^{X|\Xst}_{r-2:r-2}(\cdot | \xst_1)$ and $f^{X|\Xst}_{r-2:r-2}(\cdot | \xst_2)$ corresponding to the same eigenvalue, we can address this issue by simply finding another pair of $y$ that does not lead to this problem. In particular, 
	for a given eigenfunction $f^{X|\Xst}_{r-2:r-2}(\cdot | \xst)$,  let $D(y_1, y_2, \xst) = \left\{\tilde{\xst}: \frac{f^{X|\Xst}(y_1 | \tilde{\xst})}{f^{X|\Xst}(y_2 | \tilde{\xst})} = \frac{f^{X|\Xst}(y_1 | \xst)}{f^{X|\Xst}(y_2 | \xst)}\right\}$, the set of $\xst$ values that define eigenfunctions with the same eigenvalue. Then, any linear combination of eigenfunctions indexed by $\tilde{\xst}$ for 
	$\tilde{\xst} \in D(y_1, y_2, \xst)$ is a potential candidate for the eigenfunctions of $J_{y_1}J_{y_2}^{-1}$.  Then, define $v(\xst) \equiv \cap_{(y_1, y_2) \in \mY \times \mY} \text{span} (\{f^{X|\Xst}_{r-2:r-2}(\cdot | \tilde{\xst}): \tilde{\xst}  \in D(y_1, y_2, \xst)\})$.
	If $v(\xst) $ is one-dimensional, this set will uniquely determine the eigenfunction
	$f^{X|\Xst}_{r-2:r-2}(\cdot | \xst)$, even though such identification is up to scales. Next, we will show that if the set $v(\xst)$ has more than one dimension, then Assumption \ref{ass_distinct} would be violated. 
	When the dimension of $v(\xst)$ is greater than one, we can at least find two eigenfunctions, say
	$f^{X|\Xst}_{r-2:r-2}(\cdot | \xst)$ and $f^{X|\Xst}_{r-2:r-2}(\cdot | \tilde{\xst})$. Therefore, $\cap_{(y_1, y_2) \in \mY \times \mY} D(y_1, y_2, \xst)$ must contain at least two points $\xst$ and $\tilde{\xst}$. It follows that $\frac{f^{X|\Xst}(y_1 | \tilde{\xst})}{f^{X|\Xst}(y_2 | \tilde{\xst})} = \frac{f^{X|\Xst}(y_1 | \xst)}{f^{X|\Xst}(y_2 | \xst)}$ for any $(y_1, y_2) \in \mY \times \mY$ by definition of $D(y_1, y_2, \xst)$. Thus, Assumption \ref{ass_distinct} is violated. 
	
	Lastly, we exploit the one-to-one mapping between $f^{X|\Xst}_{r-2: r -2} (x | \Xst)$ and the parent distribution $f^{X|\Xst}(x | \Xst)$ to identify 
	$f^{X|\Xst}(x | \Xst)$ for $x \in \mA_l$.  Specifically,
	\begin{eqnarray}
		f^{X|\Xst}_{r-2:r-2}(x|\xst) &=& (r-2)  [F^{X|\Xst}(x|\xst)]^{r-3} f^{X|\Xst}(x|\xst) \notag \\
		\leftrightarrow \quad  \int^{x}_{\underline x}  f^{X|\Xst}_{r-2:r-2}(v|\xst) d v &=&  (r-2) \int^{x}_{\underline x} [F^{X|\Xst}(v)]^{r-3} f^{X|\Xst}(v|\xst) d v \notag\\
		\leftrightarrow \quad  \int^{x}_{\underline x}  f^{X|\Xst}_{r-2:r-2}(v|\xst)d v &=& [F^{X|\Xst}(x|\xst)]^{r-2} \notag\\
		\leftrightarrow F^{X|\Xst}(x|\xst) &=&   \Big[\int_{\underline x}^{x}  f^{X|\Xst}_{r-2:r-2}(v|\xst)dv \Big]^{\frac{1}{r-2}}, 
	\end{eqnarray}
	where the first equality holds by definition, the second by taking the integral over $\bar x$ to any value in the low segment. Therefore, we can identify the conditional CDF of the parent distribution at the low segment. Additionally, we can also derive the conditional marginal distribution for $x \in \mA_l$ as follows:
	\begin{eqnarray}
		f^{X|\Xst}(x|\xst) &=&  \frac{1}{r-2} \Big[\int_{\underline x}^{x}  f^{X|\Xst}_{r-2:r-2}(v|\xst)d v \Big]^{\frac{1}{r-2}-1} f^{X|\Xst}_{r-2:r-2}(x|\xst).
	\end{eqnarray}
	Note that the conditional marginal distribution in the ``low" portion is identified up to scales, ordering, and location. 
	
\paragraph{Step 2} We now proceed to identify $f^{X|\Xst}(x | \Xst)$ for $x \in \mA_h$, which is achieved given that operator $L_{X_{1: n - r + 1}|\Xst}$ is identified up to scales, ordering, and location. 

Specifically, we redefine the operator $J_y$ by abuse of notation:
	$$
	[J_{y}g](z) \equiv  \int_{\mA_l} f_{r-2,r-1,r:n}(x, y, z) g(x)dx,
	$$
	for any $y \in \mA_m$. Then, $J_y$ is a map from $\mL^2(\mA_l)$ to $\mL^2(\mA_h)$ and satisfies 
	\begin{eqnarray}
		[J_{y}g](z) &\equiv & \int_{\mathcal{X}_l} f_{r-2,r-1,r:n}(x, y, z) g(x)dx \notag \\
		&=& \int_{\mA_l} c_{r,n} \cdot  \int_{\XXst}  f^{X|\Xst}_{r-2:r-2}(x|\xst) f^{X|\Xst}(y|\xst) f^{X|\Xst}_{1:n-r+1}(z|\xst)  f^{\Xst}(\xst) d \xst g(x) dx \notag  \\
		&=&  \int_{\XXst}  f^{X|\Xst}_{1:n-r+1}(z|\xst) c_{r,n} f^{X|\Xst}(y|\xst)  f^{\Xst}(\xst) \left(\int_{\mathcal{X}_l} f^{X|\Xst}_{r-2:r-2}(x|\xst) g(x) dx\right) d \xst \notag \\
		&=& \int_{\XXst}  f^{X|\Xst}_{1:n - r + 1}(z|\xst)  c_{r,n} f^{X|\Xst}(y|\xst)  f^{\Xst}(\xst) [H_{X_{r - 2:r - 2}|\Xst}g] (\xst) d \xst \notag \\
		&\equiv & \int_{\XXst}  f^{X|\Xst}_{1: n - r + 1}(z|\xst)  [\Delta_{X=y,\Xst}H_{X_{r - 2: r - 2}|\Xst}g] (\xst) d \xst \notag \\
		&=& [L_{X_{1: n - r + 1}|\Xst} \Delta_{X=y,\Xst}H_{X_{r - 2:r - 2}|\Xst}g](z).
		\label{eq_new}
	\end{eqnarray}
	Using the same approach as Step 1 for two points $y_1$ and $y_2$ in $\mA_m$, we are able to uniquely recover the conditional distribution $f^{X|\Xst}_{1:n-r+1}(x |\xst)$ for $x \in \mA_h$, up to scales, ordering, and location. However, the ordering of UH can be matched consistently with that in Step 1, because the eigenvalues are the same in both decomposition.

Given that the conditional distribution for order statistic $X_{1: n - r + 1}$ in segment $\mA_h$ is identified up to scales, ordering, and location, we now exploit the one-to-one mapping between the parent distribution and the distribution of its order statistic to identify the parent distribution in the high segment. Specifically,
	\begin{eqnarray}
		f^{X|\Xst}_{1:n-r+1}(x|\xst) &=& (n-r+1)  [1-F^{X|\Xst}(x|\xst)]^{n-r} f^{X|\Xst}(x|\xst) \notag \\
		\leftrightarrow \quad  \int^{\bar x}_{x}  f^{X|\Xst}_{1:n-r+1}(x|\xst) d x &=& (n-r+1) \int^{\bar x}_{x}  [1-F^{X|\Xst}(x|\xst)]^{n-r} f^{X|\Xst}(x|\xst)d x \notag\\
		\leftrightarrow \quad  \int^{\bar x}_{x}  f^{X|\Xst}_{1:n-r+1}(x|\xst) d x &=&[1-F^{X|\Xst}(x|\xst)]^{n-r+1} \notag\\
		\leftrightarrow F^{X|\Xst}(x|\xst) &=& 1-  \Big[\int_{x}^{\bar x}  f^{X|\Xst}_{1:n-r+1}(x|\xst) d x \Big]^{\frac{1}{n-r+1}}.
	\end{eqnarray}
	We can then  identify the conditional marginal distribution $f^{X|\Xst}(x|\xst)$ for any $x \in \mA_h$. Note that such an identification is up to scales but has the same ordering of UH as in $\mA_l$. 
	
\paragraph{Step 3} The identification of  $f^{X|\Xst}(y|\xst) f^{\Xst}(\xst)$ for $y\in  \mathcal{X}_m$ is achieved using Equation \eqref{eq_joint3}. Specifically, since we have already identified $f^{X|\Xst}(x | \xst)$ for $x \in \mA_l \cup \mA_h$, the numerator of $\int_{y_1 \in \mA_m} f^{X|\Xst}(y_1 | \Xst) dy_1 / f^{X|\Xst}(y_2 | \Xst)  $ is known. As a result, the denominator $f^{X|\Xst}(y_2 | \xst)$ is uniquely specified for any $y_2 \in \mA_m$. Note that such identification is up to scales, ordering, and location of UH. 
	
\paragraph{Step 4}
	In summary, we can identify the conditional distribution $f^{X|\Xst}(x | \xst)$ up to scales and ordering for the three segments $ \mathcal{X}_l$,  $\mathcal{X}_m$, and $\mathcal{X}_l$. Note that the scales may vary across the three segments. We invoke the continuity of the component PDFs and the total probability argument. The scales can be pinned down using the following three restrictions.  First, the PDFs identified separately in the three segments should be the same at the cutoff points  due to the continuity of the true conditional distributions. Second, the fact that each conditional distribution should integrate to 1 provides the third restriction on the scales. These restrictions uniquely identify the scales. 
	
	Once the scales are pinned down, we exploit Assumption \ref{ass_order} to resolve the indexing problem in issue (2). Given  monotonicity, we can fix the ordering of  $\tau$ by using the conditional mean. And the further restriction on the support of UH pins down the exact location of UH. 
	

\paragraph{Step 5} The identification of the marginal distribution of UH is achieved by using the unconditional joint distribution of order statistics $X_{r-1:n}$ and $X_{r:n}$, which can be represented as
	\begin{eqnarray*}
		f_{r-1,r:n}(x,z)=\int_{\XXst} c^1_{r,n} f^{X|\Xst}_{r-1:r-1}(x|\xst)f^{X|\Xst}_{1:n-r+1}(z|\xst) f^{\Xst}(\xst) d \xst
	\end{eqnarray*}
	for any $x \leq z$. Let $K$ denote an operator mapping $g \in \mathcal{G} (\mathcal{X}_h)$ to 
	$Kg \in \mathcal{G}(\mA_l \cup \mA_m)$ with the definition:
	\begin{equation*}
		[Kg](x) \equiv  \int_{\mathcal{X}_h} f_{r-1,r:n}(x, z) g(z)dz.
	\end{equation*}
	Then, based on the above equation, we have for any $x \in \mA_l \cup \mA_m$, 
	\begin{eqnarray*}
		[Kg](x) &\equiv & \int_{\mathcal{X}_h} f_{r-1,r:n}(x, z) g(z)dz \notag \\
		&=& \int_{\mathcal{X}_h} c^1_{r,n} \cdot  \left(\int_{\XXst}  f^{X|\Xst}_{r-1:r-1}(x|\xst) f^{X|\Xst}_{1:n-r+1}(z|\xst)  f^{\Xst}(\xst) d \xst \right) g(z) dz \notag  \\
		&=&  \int_{\XXst}  f^{X|\Xst}_{r-1:r-1}(x|\xst)  c^1_{r,n}  f^{\Xst}(\xst) \left(\int_{\mathcal{X}_h} f^{X|\Xst}_{1:n-r+1}(z|\xst)g(z) dz \right) d \xst \notag \\
		&=& \int_{\XXst}  f^{X|\Xst}_{r-1:r-1}(x|\xst)  c^1_{r,n}  f^{\Xst}(\xst) [H_{X_{1:n-r+1}|\Xst}g] (\xst) d \xst \notag \\
		&\equiv & \int_{\XXst}  f^{X|\Xst}_{r-1:r-1}(x|\xst)  [\Delta_{\Xst}H_{X_{1:n-r+1}|\Xst}g] (\xst) d \xst \notag \\
		&=& [L_{X_{r-1:r-1}|\Xst} \Delta_{\Xst}H_{X_{1:n-r+1}|\Xst}g](x),
	\end{eqnarray*}
	where  the diagonal operator $[\Delta_{\Xst}g](\xst) \equiv c^1_{r,n} f^{\Xst}(\xst)g(\xst)$ for any $\xst \in \XXst$. That is to say, we obtain the following operator equivalence:
	\begin{eqnarray*}
		K = L_{X_{r-1:r-1}|\Xst} \Delta_{\Xst}H_{X_{1:n-r+1}|\Xst}.
	\end{eqnarray*}
	
	Note that operator $H_{X_{1:n-r+1}|\Xst}$ is injective and identified; operator $L_{X_{r-1:r-1}|\Xst}$ is also known and injective since we have identified the conditional density of $f^{\Xst}(y|\xst)$. The injection of $L_{X_{r-1:r-1}|\Xst}$ can be easily derived from the injection of operator $L_{X_{r-2:r-2}|\Xst}$. Hence,
	$$
	L_{X_{r-2:r-2}|\Xst}^{-1}K = \Delta_{\Xst}H_{X_{1:n-r+1}|\Xst}.
	$$
	The left side of this equation is a specified kernel, which maps a function $g \in \mG(\mA_h)$ to $\int_{\mathcal{X}_h}c^1_{r,n}  f^{\Xst}(\cdot) f^{X|\Xst}_{1:n-r+1}(z|\cdot)g(z) dz \in \mG(\XXst)$.  Based on the one-to-one mapping between the operator and its kernel, $f^{\Xst}(\xst) f^{X|\Xst}_{1:n-r+1}(z|\xst)$ is identified for any $\xst \in \XXst$. As the conditional density
	$ f^{X|\Xst}_{1:n-r+1}(z|\xst)$ has been identified previously, the marginal density of the latent factor, $f^{\Xst}(\xst)$, is then specified.  
	
	\QEDA
	\subsection{Proof of Corollary 2}
	We show that when the number of potential bidders is observed and has a large support, we can identify the conditional truncation probability. Specifically, the distribution of the number of active bidders conditional on the number of potential bidders is a mixture of the Binomial distribution with the success probability being the truncation $1-F^{X|\Xst}(R|\xst)$. Note that, the mixture weight is the marginal distribution of UH, which is already identified from the eigenfunction-decomposition. That is,
	\begin{eqnarray}
		Prob(n|N) &=&\int_{\xst \in \XXst }Prob(n|N,\xst) d F^{\Xst}(\xst) \notag \\
		&=& \int_{\xst \in \XXst} C_{N,n} (1-F^{X|\Xst}(R|\xst))^n (F^{X|\Xst}(R|\xst))^{N-n}d F^{\Xst}(\xst),
	\end{eqnarray}
	where $Prob(n|N)$ is estimable from the data, $C_{N,n}$ is a constant,  $F^{\Xst}(\xst)$ is identified, and $F^{X|\Xst}(R|\xst)$ is the unknown and the object of interest, which is continuous because the unobserved heterogeneity $\xst$ is continuous. 
	
	It is worth emphasizing that our goal is to identify the UH-specific truncated probability $F^{X|\Xst}(R|\xst)$ given that the mixture distribution $F^{\Xst}(\xst)$ can be viewed as known. In contrast, the identification results developed in the mixture literature mainly focus on identifying the mixture distribution of the success probability, which takes any value in the interval $[0,1]$. That is, let $p$ denote a general success probability, where $p \in [0,1]$, and $G(p)$ denote the cumulative distribution of $p$, and the goal is to identify $G(p)$ for any $p$. 
	
	We show that our identification problem can be viewed as the dual problem of that in the conventional mixture model. Specifically, define a random variable $P$, which is a function of the unobserved heterogeneity $\Xst$ using the truncated probability, $P\equiv 1-F^{X|\Xst}(R|\Xst)$. Suppose the conditional truncating probability varies with UH $\xst$. That is, for any $p \in [0,1]$, there is at most one $\xst$ such that  $p=1-F^{X|\Xst}(R|\xst)$.  We can then  reverse this relationship and represent the unobserved $\xst$ as a function of the success probability $p$, i.e., $\xst=F^{-1}_{X|\Xst}(1-p,R)$. Therefore, we can reformulate the above identification problem to resemble the problem in the mixture literature and thus can apply their results directly. That is,
	\begin{eqnarray}
		Prob(n|N) &=& \int_{p \in [0,1]} C_{N,n} p^n (1-p)^{N-n}d F^{\Xst}(F^{-1}_{X|\Xst}(1-p,R)) \notag \\
		&=& \int_{p \in [0,1]} C_{N,n} p^n (1-p)^{N-n}d G(p),
	\end{eqnarray}
	where the first equality holds by change of variables $\xst=F^{-1}_{X|\Xst}(1-p,R)$ and $p=1-F^{X|\Xst} (R|\xst)$, and the second equality holds by redefining $G(p)=F^{\Xst}(F^{-1}_{X|\Xst}(1-p,R))$, which is the cumulative distribution of $p$. Therefor, for some $p \in [0, 1]$ that there does not exist an unobserved value $\tau$ such that $\xst=F^{-1}_{X|\Xst}(1-p,R)$, $G(p)=0$. Note that $G(p)$ is unknown because $F^{X|\Xst}(R|\Xst)$ is unknown and $F^{-1}_{X|\Xst}(1-p,R)$ is unknown. 
	
	We then show that the first $N$ moments of the mixture distribution $G(p)$ is identifiable from the distribution of the active bidders conditional on the number of potential bidders. The conditional distribution provides information on the first $N$ moments. Specifically, 
	\begin{eqnarray}
		Prob(n|N) &=& E [C_{N,n} p^n (1-p)^{N-n}] \notag \\
		&=& E[ C_{N,n} p^n  \sum^{N-n}_{r=0} C_{N-n, r} (-1)^{r} p^r] \notag \\
		&=&  \sum^{N-n}_{r=0} C_{N-n, r}C_{N,n} (-1)^{r} E [ p^{r+n}] \notag \\
		&=&\sum^{N}_{i=n} C_{N-n, i-n}C_{N,n} (-1)^{i-n} E [ p^{i}] \notag \\
		&=&\sum^{N}_{i=1} C_{N-n, i-n}C_{N,n} (-1)^{i-n} E [ p^{i}],
	\end{eqnarray}
	where the first equality is by change of variables, the second equality holds by expressing $(1-p)^{N-n}$, the fourth equality holds by letting $i=r+n$, and the last equality holds by $C_{N-n, i-n}=0$ for $i-n\le 0$. We can then  rewrite the above equation in matrix form
	\begin{eqnarray}
		P_{N}=Am_{N},
	\end{eqnarray}
	where $P_{N} \equiv \{Prob(n=1|N),...,Prob(n=N|N)\}^T$ is an $N\times 1$ column vector, $A \equiv \{C_{N-n, i-n}C_{N,n} (-1)^{i-n}\}_{i,n}$ is a $N \times N$ square matrix, and $m_{N} \equiv \{E [ p], E [ p^{2}],...,E [ p^{N} ]\}^T$ is a $N \times 1$ column vector collecting the $N$ moments. Note that $A$ is invertible because it is an upper triangular matrix with nonzero diagonal elements. Therefore, we can identify the first $N$ moments of distribution $G(p)$. 
	
	If $N$ has a large support, following the exiting literature \citep{gut2005probability} that the distribution of a bounded variable is uniquely determined by its moments, we can identify all moments of $G(p)$ so that $G(p)$ is identified. Consequently, the truncated distribution $1-F^{X|\Xst}(R|\xst)$ is identified. Specifically, for a given $p$, we have $G(p)=F^{\Xst}(\xst)$, so that we can recover $\tau$ by inverting the CDF of $F^{\Xst}$ since it is identified. Moreover, given that $p=1-F^{X|\Xst}(R|\tau)$, we can have $F^{X|\Xst}(R|\tau)=1-p$. Therefore, we can recover the value of the  truncation distribution $F^{X|\Xst}(R|\tau)$ for any given $\tau$.
	
	\QEDA
	\section{Properties of the Sieve Estimators} \label{sec:properties}
	
	
We provide sieve estimators to approximate the joint distributions of bids and unobserved heterogeneity. This is different from the estimator proposed in \cite{hu2008instrumental}, where they approximate the conditional distribution of bids and the marginal distribution of the unobserved heterogeneity. Our sieve estimator for approximating the joint distribution has the following advantages:  First, HS concerns the relationship between the dependent variable and the independent variable, and the relationship between the latent variable and the instrumental variable. In particular, they consider a parametric model for the conditional density of the dependent variable given the latent variable and nonparametric models for the conditional density of the independent variable and that of the latent variable. For the latter two, they need to consider different sieve estimates. Consequently, for each value of the latent variable and the instrument, normalization is needed, which is more computationally burdensome. In contrast, we are concerned about the bid/value distribution given the latent variable. The order statistics of bids follow the same parent distribution. Hence, we can first use a bivariate sieve to approximate the joint density of the bid and the latent variable. The densities of interest can then be directly derived from the joint sieve estimator. Only one normalization step is needed when implementing our sieve estimation procedure. Second, in \cite{hu2008instrumental}, the density of the dependent variable given the latent is assumed to be parametric, while the conditional densities of the independent variable and the latent variable are modeled in a nonparametric way. Model misspecification may be encountered due to the parametric assumption. However, our sieve estimator is under a fully nonparametric framework, and thus we can circumvent the issue of model misspecifications. In practice, concerns about slow convergence might be raised because of this fully nonparametric structure. However, numerical studies demonstrate the resulting estimator from the sieve estimator has a promising performance. Lastly, we adopt a new technique to derive the consistency of our estimator. More specifically, we establish a concentration inequality based on the bracketing entropy and, from there, we derive the consistency of our sieve estimator, while 
\cite{hu2008instrumental} uses a covering number. 	
	
	\subsection{Asymptotic Properties}

	
	\noindent{\it Proof of Lemma \ref{lema_inequa}:} 		
	Since $(\hat{f}_m^{X | \Xst}, \hat{f}_m^{\Xst})$ maximizes $\bP_m[\log G(x, y, z; f)]$ over the sieve space, where $\bP_m$ denotes the empirical measure of the data $(x_i, y_i, z_i)_{i = 1}^m$,  it follows that
	$$
	\bP_m\left\{\log G(x, y, z; \hat{f}_m^{X | \Xst}, \hat{f}_m^{\Xst})\right\} \geq \bP_m\left\{\log G(x, y, z; f_m^{X | \Xst}, f_m^{\Xst})\right\}.
	$$
	Therefore, we have
	\begin{align}
		\nonumber
		m^{-1/2} \bG_m\left\{\log \frac{G(x, y, z; \hat{f}_m^{X | \Xst}, \hat{f}_m^{\Xst})}{G(x, y, z; f_m^{X | \Xst}, f_m^{\Xst})}  \right\} & \geq \bP\left\{\log \frac{G(x, y, z; f_m^{X | \Xst}, f_m^{\Xst})}{G(x, y, z; f_0^{X | \Xst}, f_0^{\Xst})}  \right\} \\
		& ~~+ \bP\left\{\log \frac{G(x, y, z; f_0^{X | \Xst}, f_0^{\Xst})}{G(x, y, z; \hat{f}_m^{X | \Xst}, \hat{f}_m^{\Xst})}  \right\}. 
		\label{ineq-mle}
	\end{align}

	\QEDA
	
	\bigskip

	\noindent{\it Proof of Theorem \ref{consistent}}: 
	Under Assumption \ref{ass_appro}, to find the convergence rate of the sieve MLE, we only need to quantify the convergence rate of the estimation error in the sieve space.  
	
	Under Assumption \ref{ass-bracket}, by Theorem 19.35 of \cite{van1998}, we have
	$$
	\sqrt{m}\|\bP_m - \bP\|_{\mF_m} \leq O_p(1) p_m Q_m^{c_0} \log p_m/ \sqrt{m} 
	$$
	for some positive constant $c_0$. Let (I) and (II) denote the two terms of the right-hand side of \eqref{ineq-mle}, respectively. Since the functional $G$ is Lipschitz continuous with respect to each component, we have that
	$$
	(I) \geq -O_p(1)\left\{\|f_m^{X | \Xst} - f_0^{X | \Xst}\|_{L_{\infty}} + \|f_m^{\Xst} - f_0^{\Xst}\|_{L_{\infty}}\right\} \geq -O_p(p_m^{-\beta}). 
	$$
	Note that (II) is the Kulback-Leibler information. We consider the Taylor expansion of it. Obviously, the first term in this expansion vanishes while the second-order term in the expansion has a lower bound,
	$$
	O(e^{-c_1Q_m})\big\|G(x, y, z; f_0^{X | \Xst}, f_0^{\Xst}) - G(x, y, z; \hat{f}_m^{X | \Xst}, \hat{f}_m^{\Xst}))\big\|_{L_2(P)}^2,
	$$
	for some positive constant $c_1 > 1$. 
	
	Since we assume that the joint density is bounded, the joint probability measure $P$ is equivalent to the product of the Lebsgue measure in $[0, 1]^3$ .
	Therefore, by combining the results above, we have 
	\begin{equation*}
		\int_{0 \leq x \leq y \leq z \leq 1} \left\{G(x, y, z; f_0^{X | \Xst}, f_0^{\Xst}) - G(x, y, z; \hat{f}_m^{X | \Xst}, \hat{f}_m^{\Xst})\right\}^2 dx dy dz \leq O_p(1)B(m, p_m, Q_m),
	\end{equation*}
	where $B(m, K_m, Q_m) = e^{c_2Q_m} p_m \log p_m / \sqrt{m} + e^{c_2Q_m} / p_m^{\beta}$, and $c_2 > 1$ is a constant.

	Lastly, we show the consistency of $\hat{f}_m^{X | \Xst}$ and  $\hat{f}_m^{\Xst}$. To do this, we can consider the square $L_2$ distance between
	$\int_{0 \leq x \leq y \leq z \leq 1} G(x, y, z; f_0^{X | \Xst}, f_0^{\Xst}) dx dz$ and \\
	$\int_{0 \leq x \leq y \leq z \leq 1}G(x, y, z; \hat{f}_m^{X | \Xst}, \hat{f}_m^{\Xst})]^2 dx dz$.  This upper bound still holds for this distance. After some simple algebra,
	it follows that
	$$
	\int \left\{\hat{f}_m^{X | \Xst} (y | \xst)  \hat{f}_m^{\Xst}(\xst)   - f_0^{X | \Xst} (y | \xst)  f_0^{\Xst}(\xst)\right\}^2 d\xst \leq O_p(1)B(m, p_m, Q_m).
	$$
	
	We can further justify that $\|\hat{f}_m^{X, \Xst} - f_0^{X, \Xst}\|^2_{L^2}$ is bounded by $O_p(1)B(m, p_m, Q_m)$. Then, it is obvious that $B(m, p_m, Q_m)^{1/2}$ is the convergence rate of 
	both $\hat{f}_m^{X | \Xst} (x | \xst)$ and $\hat{f}_m^{\Xst} (\xst) $. 		
	\QEDA


	\subsection{Popular Sieve Spaces in Auctions}
	
	
	In this subsection, we evaluate the regularity assumptions for B-splines and Bernstein polynomials, which are popular in empirical applications due to their flexibility and the ease with which shape can be imposed. See, e.g., \citet{compiani2018common} and \citet{kong2020not} for using Bernstein polynomials in auctions. 
	
	\subsubsection{B-Splines}
	
	We want to show that the proposed MLE using B-splines as a sieve base satisfies the regularity conditions in Theorem \ref{consistent}. We first express the joint density as
	$$
	f^{X, \Xst}(x, \xst) =  \frac{\exp\{\eta(x, \xst)\}}{\int\int \exp\{\eta(x, \xst)\} dxd\xst},
	$$
	which is always positive and integrates to 1, qualifying it as a density function. As a result, the conditional and marginal densities can be expressed as $f^{X | \Xst}(x | \xst) = \frac{\exp\{\eta(x, \xst)\}}{\int \exp\{\eta(x, \xst)\} dx}$ and $f^{\Xst}(\xst) = \frac{\int_x\exp\{\eta(x, \xst)\}dx}{\int\int \exp\{\eta(x, \xst)\} dxd\xst}$, respectively.
	
	Next, we employ a tensor product of B-spline basis functions to approximate $\eta(x, \xst)$. 
	We first define an extended partition on the interval [0, 1], given by
	$$
	\triangle_e = \{s_{-L+1} = \cdots = s_{-1} = 0 = s_0 < s_1 < \cdots < s_{K_m + 1} = 1 = \cdots = s_{L + K_n}\},
	$$
	where $L$ is the order of the spline basis and $K_m$ is the number of interior knots. Let $\{B_j^L(t)\}_{j = 1}^{K_m + L}$ be a normalized B spline basis of order
	$L$ (degree $L - 1$) associated with $\triangle_e$. The sieve space for the parameter $\eta(x, \xst)$ is defined as
	\begin{align} 
		\nonumber
		& S_m(L, K_m, Q_m)  = \\
			 &\left\{\eta(x, \xst): \sum_{i_1, i_2 = 1}^{K_m + L} |b_{i_1, i_2}| \leq Q_m,  \eta(x, \xst) =   \sum_{i_1, i_2 = 1}^{K_m + L} b_{i_1, i_2} B_{i_1}^L (x) B_{i_2}^L(\xst), \sum_{i_1, i_2 = 1}^{K_m + L} b_{i_1, i_2} B_{i_1}^L (0) = 0\right\}. \label{eq-sieve}	
	\end{align}
	Here, $Q_m$ is a constant that depends on sample size $m$.  The first two conditions in $S_m(L, K_m, Q_m) $ ensure that the sieve space is a compact set in a finite-dimensional space, and the third condition is equivalent to $\eta(0, \xst) = 0$, which is needed to ensure the identifiability of $\eta$.

	We now prove that the sieve estimator using B-splines as a sieve base is consistent. Specifically, we show that the sieve base of B-splines satisfies Assumption \ref{ass_appro}, which is critical for consistency. We introduce the following two regularity conditions:
	\begin{itemize}
		\item (A1)  
		For a known integer $k \geq 2$, the true conditional density $f_0(x, z)$ satisfies $\log f_0(x, \xst) \in W^{k, \infty} ([0, 1]^2)$, where $ W^{k, \infty} ([0, 1]^2)$ is a Sobolev space consisting of the functions defined on $[0, 1]^2$ with bounded $k$th derivative.
		\item (A2) $(Q_m, p_m)$ satisfies $Q_m = O(\log \log m)$ and $p_m = O(m^{\alpha})$ with $0 < \alpha < 1/4$.
		
	\end{itemize}
	
	Assumption A1 characterizes the smoothness of the joint density function; Assumption A2  is equivalent to Assumption \ref{ass:bound}. Together with Assumption \ref{ass_comp}, these assumptions ensure that the logarithm of the true joint density function can be approximated sufficiently well by the tensor product of B-spline basis functions, i.e., Assumption \ref{ass_appro} is satisfied. This is also one important reason why we prefer choosing B-spline  as the univariate basis functions to construct the sieve space. We summarize this result in the following proposition.

	\begin{lemma} \label{prop_Bspline}
		Under Assumptions A1 and \ref{ass_comp}, Assumption \ref{ass_appro} is satisfied. That is, there exist $f_m^{X | \Xst}(x | \xst)$ and $f_m^{\Xst}(\xst)$, both of which are represented in terms of basis functions in the sieve space, such that 			
		\begin{align*} 			 
			\|f_m^{X | \Xst}(x | \xst) - f_0^{X | \Xst}(x | \xst)\|_{L_{\infty}([0, 1]^2)} &= O(p_m^{-k}), \\
			\|f_m ^{\Xst}(\xst) - f_0^{\Xst} (\xst)\|_{L_{\infty}([0, 1]^2)} & = O(p_m^{-k}).
		\end{align*}
	\end{lemma}
	
	\noindent{\it Proof of Lemma \ref{prop_Bspline}:} 
	Following \cite{sch1981}, we define a linear operator $\mathcal{Q}_p$, which is a mapping from $W^{k, \infty} ([0, 1]^p)$ to the sieve space. 
	More specifically,
	for any $g \in W^{k, \infty} ([0, 1]^p)$,
	$$
	\mathcal{Q}_p[g] = \sum_{i_1, \ldots, i_p = 1}^{K_m + L} \Gamma_{i_1, \ldots, i_p}[g]  B_{i_1}^L (x_1)\cdots B_{i_p}^L(x_p),
	$$
	where $\Gamma_{i_1, \ldots, i_p}$ are the linear functionals in $L_{\infty}([0, 1]^p)$.  This mapping satisfies that
	$$
	\sum_{i_1, \ldots, i_p = 1}^{K_m + L} |\Gamma_{i_1, \ldots, i_p}[g]| \leq (2L + 1)^p 9^{p(L - 1)} ||g||_{L_{\infty}([0, 1]^p)},
	$$
	and by Theorem 12.7 of \cite{sch1981},
	$$
	\|\mathcal{Q}_p[g] - g \|_{L_{\infty}([0, 1]^p)} \leq \frac{C(L)}{K_m^k} \|g\|_{W^{k, \infty} ([0, 1]^p)}. 
	$$
	
	Then, we define $\eta_{m}(x, \xst) = \mathcal{Q}_2 [\log f_0 (x, \xst)] - \mathcal{Q}_2 [\log f_0 (x, \xst)] |_{x = 0}$, an element in the sieve space. Hence,
	$$
	f_m^{X | \Xst}(x | \xst) = \frac{\exp\{\eta_m(x, \xst)\}} {\int_0^1 \exp\{\eta_n(x, \xst)\} dx},~~~f_m^{\Xst}(\xst) = \frac{\int_0^1 \exp\{\eta_m(x, \xst)\} dx}{\int_0^1\int_0^1 \exp\{\eta_m(x, \xst)\} dxd\xst}.
	$$
	As a result, 
	$$
	\|f_m^{X | \Xst}(x | \xst) - f_0^{X | \Xst}(x | \xst)\|_{L_{\infty}([0, 1]^2)} \leq O(1) \| \log f_0 - \mathcal{Q}_2 [\log f_0] \|_{L_{\infty}([0, 1]^2)} \leq O(K_m^{-k}).
	$$
	Moreover, the same bound holds for 
	$\|f_m - f_0\|_{L_{\infty}([0, 1]^2)}$ and $\|f_m^{\Xst} - f_0^{\Xst}\|_{L_{\infty}([0, 1]^2)}$.
	\QEDA
	
	We then prove that Assumption \ref{ass-bracket} is satisfied for sieve MLE using B-splines.  If the sieve space is $\mS_m$, constructed as in Equation \ref{eq-sieve}, and Assumption \ref{ass_comp} and conditions A1-A2 are met, then the following space
	\begin{equation*}
		\begin{split}
			\left\{\log \frac{G(x, y, z; \tilde{f}_m^{X | \Xst}, \tilde{f}_m^{\Xst})}{G(x, y, z; f_m^{X | \Xst}, f_m^{\Xst})}:  \tilde{f}_m^{X | \Xst} = \frac{\exp\{\tilde{\eta}_m(x, \xst)\}} {\int_0^1 \exp\{\eta_m(x, \xst)\} dx}, \right.\\
			\left. \tilde{f}_m^{\Xst} = \frac{\int_0^1 \exp\{\tilde{\eta}_m(x, \xst)\} dx}{\int_0^1\int_0^1 \exp\{\tilde{\eta}_m(x, \xst)\} dxd\xst}, \tilde{\eta}_m \in S_m \right\}
		\end{split}
	\end{equation*}
	satisfies Assumption \ref{ass-bracket}. A similar result can be found in \cite{zeng2005}.

	\begin{corollary}
		Under Assumption \ref{ass_comp} and the regularity conditions (A1) and (A2), the proposed MLE using B-splines as sieve bases is consistent.
	\end{corollary}

	\subsubsection{Berstein Polynomials}
	
	We approximate the joint density function of $\Xst$ and $X$ using a mixture of beta distributions as follows: 
	\[
	f(x,x^*) = \sum_{i,j} \theta_{ij} \beta_i(x) \beta_j(x^*) , 
	\]
	where $\beta_i(x)$ represents a beta density function with parameters $\alpha=i,\beta=p_m+1-i$, and $p_m$ is the number of components in the mixture.\footnote{One can allow different numbers for each of the two dimensions. We leave the optimal choice for future research.} We restrict $\theta_{ij}\geq 0$ and $\sum_{i,j\in \{1,\ldots,p_m\}} \theta_{ij} = 1$ to ensure that $f$ is a joint density function. That is, $$
	\int_{x,x^*} f(x,x^*) dx dx^* = \sum_{i,j\in \{1,\ldots,p_m\}} \theta_{ij} = 1.$$ 
	Given the approximation of the joint distribution, we can represent the marginal density function of $x^*$ as 
	\[
	f^{\Xst}(x^*) = \int_x f(x,x^*) dx = \int_x \sum_{i,j} \theta_{ij} \beta_i(x) \beta_j(\xst) dx = 
	\sum_j \left\{\left(\sum_{i} \theta_{ij}\right) \beta_j(x^*)\right\},\]
	the conditional probability density function as
	\[
	f^{X|\Xst}(x|x^*) = \frac{f(x,x^*)}{f^{\Xst}(x^*)}  = \frac{\sum_{i,j} \theta_{ij} \beta_i(x) \beta_j(x^*)}{\sum_j [(\sum_{i} \theta_{ij}) \beta_j(x^*)]},
	\]
	and the conditional cumulative distribution function as
	\[
	F^{X|\Xst}(x|x^*)  = \int_{-\infty}^x f^{X|\Xst}(x|x^*) dx = \frac{\sum_{i,j} \theta_{ij} B_i(x) \beta_j(x^*)}{\sum_j [(\sum_{i} \theta_{ij}) \beta_j(x^*)]},
	\]
	where $B_i(x)$ represents the CDF of the beta distribution $\beta_i(x)$.
	
	According to these derivations, given $\Xst = \xst$, we can write the density functions as mixtures of Beta distributions:
	\begin{align*}
		f^{\Xst}(\xst) = \sum_{j = 1}^{p_m} w_{1j} \beta_j(\xst), \quad
		f^{X | \Xst}(x | \xst) =  \sum_{i = 1}^{p_m} w_{2i}(\xst) \beta_i(x)
	\end{align*}
	for some nonnegative weights $\{w_{1j}, j = 1, \ldots, p_m\}$ and $\{w_{2i}(\xst), i = 1, \ldots, p_m\}$ that satisfy $\sum_{j = 1}^{p_m}w_{1j} = 1$ and $\sum_{i = 1}^{p_m}w_{2i}(\xst) = 1$. Let $\hat{f}_m^{X | \Xst} (x | \xst)$ and $\hat{f}_m^{\Xst} (\xst)$ denote the sieve MLE of $f_0^{X | \Xst}(x | \xst)$ and $f_0^{\Xst}$, respectively. 
	
	The following result is from Theorem 5 of \cite{petrone}. 
	If we assume that $p_m = m^{\alpha}$ for some $\alpha \in (0, 1)$, and the true joint density $f_0(x, \xst)$ is continuous and bounded away from 0, then we have for every $\epsilon > 0$, 
	\begin{align*}
		P\{d_H(f_0^{X | \Xst}(x | \xst),  \hat{f}_m^{X | \Xst} (x | \xst)) > \epsilon\} < 4\exp(-mc_1\epsilon^2), \\ P\{d_H(f_0^{\Xst}(\xst),  \hat{f}_m^{ \Xst} (\xst)) > \epsilon\} < 4\exp(-mc_2\epsilon^2)
	\end{align*}
	for some positive constants $c_1$ and $c_2$, where $d_H(f, g) = \left[\int_0^1 \{\sqrt{f(x)} - \sqrt{g(x)}\}^2 dx\right]^{1/2}$ is the Hellinger metric between two density functions with  domain $[0, 1]$. It follows that both $\hat{f}_m^{X | \Xst} (x | \xst))$ and $\hat{f}_m^{ \Xst} (\xst)$ are consistent. A key technique to obtain this result is to make use of the fact that, when the sieve space $\mathcal{F}_m$ is taken as the mixture of Beta distributions with $p_m$ components, then $N_{[]}(\epsilon, \mathcal{F}_m, d_H) = O((1/\epsilon)^{p_m})$. See Proposition 2 of \cite{petrone}. Thus, Assumption \ref{ass-bracket} is satisfied for the sieve space comprised of Beta mixtures.
	
	Additionally, as stated in \cite{ghosal2001}, if $g(x)$ is a continuously differentiable
	probability density on (0,1] with bounded second derivative, then there exists a beta mixture density, such that
	$$
	\sup_{0 < x \leq 1}\bigg |g(x) - \sum_{j = 1}^{p_m}\theta_j \beta_j(x) \bigg| = O(p_m^{-1}). 
	$$
	Moreover, if $\sum_{j = 1}^{p_m} \theta_j = 1$ with each $\theta_j \in [0, 1]$, then 
		$$
	\sup_{0 < x \leq 1}\bigg |\sum_{j = 1}^{p_m}\theta_j \beta_j(x) \bigg| \leq p_m. 
	$$
	Therefore, if $\mathcal B_m$ is constructed using a mixture of Beta distributions, and the true joint density $f_0^{X, \Xst}$ is continuously differentiable and has bounded second derivatives, Assumptions \ref{ass_appro}-\ref{ass-bracket} are satisfied with suitable conditions on $p_m$ and $Q_m$, as in Assumption \ref{ass:bound}.
	
	\begin{corollary}
		Assume that the true joint density $f_0^{X, \Xst}$ is continuously differentiable and has bounded second derivatives. Under Assumption \ref{ass_comp} and suitable conditions on $p_m$ and $Q_m$, the proposed MLE using Berstein polynomials as a sieve base is consistent.
	\end{corollary}

	\subsection{Estimation with Bernstein Polynomials: Finite Sample Property} \label{bern}
	
	To demonstrate the finite sample property of the sieve estimator using Bernstein polynomials, we conduct the following Monte Carlo experiments. The results show that the sieve estimators perform well even with modest sample sizes. 
	
	The data generating process is as follows. We first generate the latent variable $X^*$ using beta distribution $Beta(\alpha^*,\beta^*)$, where $\alpha^* = 3$ and $\beta^* = 1.5$. We then generate the measurement $X$ using the conditional distribution that is specified again as the beta distribution, with $\alpha(x^*) = 1.5$ and $ \beta(x^*) =1.5 (1+x^*)$. To generate a set of order statistics, we generate $4$ measurements for each $x^*_t$ and record the lowest three, i.e., $X_{1:n}=x^t, X_{2:n}=y^t, X_{3:n}=z^t$. Repeating this process $1,000$ times produces a sample of $1,000$ consisting of the lowest three order statistics.

\begin{figure}[ht!]
		\centering
		\includegraphics[width=0.75\textwidth]{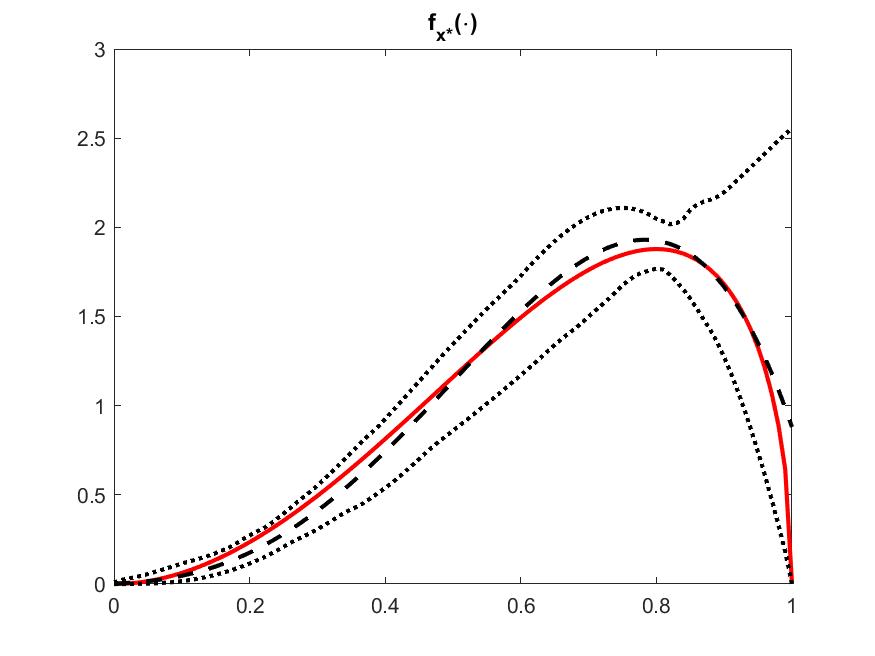}
		\caption{Estimation with Bernstein Polynomials: $f_{x^*}$}
		\label{fig:xstar}
	\end{figure}

\begin{figure}[ht!]
		\centering
		\includegraphics[width=0.75\textwidth]{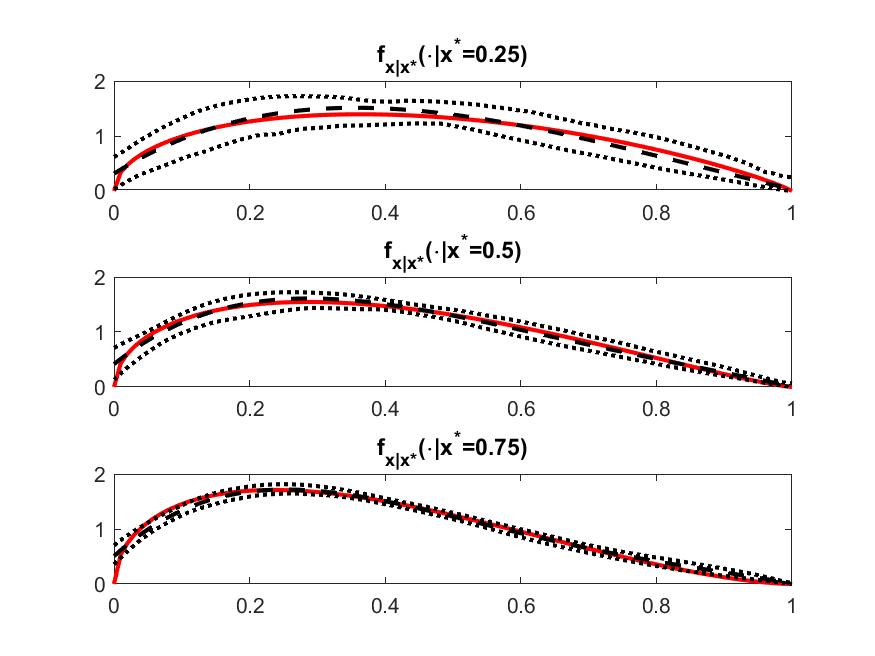}
		\caption{Estimation with Bernstein Polynomials: $f_{x|x^*}$}
		\label{fig:x}
	\end{figure}

	In estimation, we use $L=5$. To impose the restrictions that $\theta_{ij} \geq 0, \sum_{ij} \theta_{ij}=1$, we re-parametrize $\theta_{ij} = \frac{\exp(\gamma_{ij})}{\sum_{ij}\exp(\gamma_{ij})}$, where $\gamma_{11}$ is normalized to be 0. We replicate the estimation 100 times and report the 5\%, mean, and 95\% quantile of the estimated marginal distribution $f^{\Xst}(x^*)$ and the conditional distributions $f^{X|\Xst}(x|x^*)$ for $x^* =0.25, 0.50, 0.75$. See Figures \ref{fig:xstar} and \ref{fig:x}, where red solid lines represent the true densities, black dash lines represent the mean of the estimates, and black dotted lines represent the 5\% and 95\% quantiles.
	
	\subsection{Some Sieve Examples} 
	For reference, we provide a few alternative sieve bases: 1) Trigonometric linear series as the base, where the space of Trigonometric polynomials on the real line of degree $L$ or less can be represented as:
	\[\mbox{TriPol}(L)= \left\{a_0, \sum^{L}_{k=1} [a_k \cos(2k\pi x)+b_k \sin(2k\pi x)], x \in [0,1]; a_k,b_k \in \Re \right\};\]
	2) Hermite polynomials as the base, where the space of Hermite polynomials on the real line of degree $L$ or less is represented as:
	\[\mbox{HPol}(L) = \left\{\sum^{L+1}_{k = 1} a_k H_k(x) \exp\left(-\frac{x^2}{2}\right), x \in \Re: a_k \in \Re)\right\},\]
	where $H_k(x)$ is the probabilists' Hermite polynomials.

	\section{Identification using Four Order Statistics} \label{sec:four}
	
	Here, we extend the main identification results to the situation with four order statistics that are not necessarily consecutive. The identification is also conducted on different segments of the support due to the special feature of order statistics.  The identification process consists of several steps: 1) An eigenfunction-decomposition argument identifies a linear integral operator, which is defined using both the conditional distribution on a higher order statistic and the UH as kernel function on the low segment. Therefore, we can identify the conditional distribution in such a segment. 2) We use the fact that the operator defined on the low segment is identified to identify the operator defined on the high segment and so identify the conditional distributions. 3) Both identifications are up to scales, ordering, and location. We exploit similar features as in Theorem 1 to solve these problems. 
	
	Suppose we observe four order statistics that are not necessary consecutive. Denote the four order statistics as $r_i$, where $i=1, \ldots, 4$ and $1\le r_1<r_2<r_3<r_4 \le n$. For simplicity of notation, let $x_i$ denote the realized values of order statistics $X_{r_i:n}$ and omit $n$ in all notation. The joint distribution of the observed four order statistics can be represented as
	\begin{align*}
		&\quad f_{r_1,r_2,r_3,r_4:n}(x_{1}, x_{2},x_{3},x_{4})\notag \\
		&= \int_{\xst} f_{r_1,r_2,r_3,r_4:n}(x_{1}, x_{2},x_{3},x_{4}|\xst) f^{\Xst}(\xst) d \xst \notag\\
		&=\mathbbm{1}(x_1 \le x_2 \le x_3 \le x_4)  \cdot \int_{\xst} f_{r_1|r_2:n}(x_{1}|x_2,\xst)   f_{r_4|r_3:n}(x_{4}|x_3,\xst) f_{r_2,r_3:n}(x_{2},x_3|\xst) f^{\Xst}(\xst) d \xst.\notag
	\end{align*}
	The last equality holds due to the Markov property in order statistics. This joint distribution of four order statistics has a semi-separable structure, in the sense that we can separate the observed joint density function into three density functions. This, again, is similar to that in the measurement error literature, but it has an extra restriction by the nature of order statistics $\mathbbm{1}(x_1 \le x_2 \le x_3 \le x_4)$, which cannot be separated but can be controlled by dividing the support accordingly. 
	
	Once the correlation between order statistics is controlled by restricting the variation of the order statistics into the associated sub-support, we  can derive the equivalence between the linear operator defined in the data and the unknown densities that we are yet to identify. Given that the injective and distinct assumptions are satisfied, we can then identify the latent density uniquely. Please refer to the identification details in the online supplement.

	To control for the correlation between order statistics, we follow the identification argument in the situation with three consecutive order statistics. Specifically, we divide the support into four segments and only exploit the variations of $x$ in the predetermined segments: $x_1 \in \mathcal{X}_l \equiv \{x: x \le c_1\}, x_2 \in \mathcal{X}_{m1} \equiv [c_1,c_2], x_3 \in \mathcal{X}_{m2}\equiv [c_2, c_3], x_4 \in \mathcal{X}_h \equiv \{x: x \ge c_2\}$. The separable structure of the joint distribution $f_{r_1,r_2,r_3,r_4:n}(x_{1}, x_{2},x_{3},x_{4})$ reappears then. Specifically, if $x_1 \in \mathcal{X}_l, x_2 \in \mathcal{X}_{m1}, x_3 \in \mathcal{X}_{m2}, x_4 \in \mathcal{X}_h$, the joint distribution can be expressed as
	\begin{eqnarray*}
&&f_{r_1,r_2,r_3,r_4:n}(x_{1}, x_{2},x_{3},x_{4}) \notag \\
&=&\int_{\xst} f_{r_1|r_2:n}(x_{1}|x_2,\xst)   f_{r_4|r_3:n}(x_{4}|x_3,\xst) f_{r_2,r_3:n}(x_{2},x_3|\xst)f^{\Xst}(\xst) d \xst. 
	\end{eqnarray*}
	
\paragraph{Step 1} We then exploit the equivalence of linear integral operators to identify the conditional distribution. Particularly, we can derive the following operator equivalence, fixing $x_2 \in \mathcal{X}_{m1}$ and $x_3 \in \mathcal{X}_{m2}$:
	\begin{eqnarray}
		&&	[J_{x_2,x_3}g](x_1) \notag \\
		&\equiv & \int_{x_4\in \mathcal{X}_h} f_{r_1,r_2,r_3,r_4:n}(x_{1}, x_{2},x_{3},x_{4}) g(x_4)dx_4 \notag \\
		&=& \int_{x_4 \in \mathcal{X}_h}  \int_{\xst}  f_{r_1|r_2:n}(x_{1}|x_2,\xst)   f_{r_4|r_3:n}(x_{4}|x_3,\xst) f_{r_2,r_3:n}(x_{2},x_3|\xst) f^{\Xst}(\xst) g(x_4) dx_4 \notag  \\
		&=&  \int_{\xst} f_{r_1|r_2:n}(x_{1}|x_2,\xst)   f_{r_2,r_3:n}(x_{2},x_3|\xst) f^{\Xst}(\xst) \int_{x_4 \in \mathcal{X}_h}  f_{r_4|r_3}(x_{4}|x_3,\xst)g(x_4) dx_4 d \xst \notag \\
		&=&  \int_{\xst} f_{r_1|r_2:n}(x_{1}|x_2,\xst)   f_{r_2,r_3:n}(x_{2},x_3|\xst) f^{\Xst}(\xst) [H_{X_{r_4}|X_{r_3}=x_3,\Xst} g](x_4) \xst \notag \\
		&\equiv & \int_{\xst} f_{r_1|r_2:n}(x_{1}|x_2,\xst)   [\Delta_{x_{2},x_3,\Xst} H_{X_{r_4}|X_{r_3}=x_3,\Xst} g] (\xst) d \xst \notag \\
		&=& [L_{X_{r_1}|x_2,\Xst} \Delta_{x_{2},x_3,\Xst}H_{X_{r_4}|x_3,\Xst}g](x_1).
		\label{eq_joint21}
	\end{eqnarray}
	Equation \eqref{eq_joint21} implies that the operators from both sides are equivalent for any $x_2 \in \mathcal{X}_{m1}, x_3 \in \mathcal{X}_{m2}$. That is, 
	\begin{eqnarray}
		J_{x_2,x_3}= L_{X_{r_1}|x_2,\Xst} \Delta_{x_{2},x_3,\Xst}H_{X_{r_4}|x_3,\Xst}.
		\label{eq_joint23}
	\end{eqnarray}
	Since such equivalence holds for any $x_2 \in \mathcal{X}_{m1}, x_3 \in \mathcal{X}_{m2}$, we first have the following equations at two different values of $(X_{r_2}, X_{r_3})$: $(c_1,x_3)$ and $(x_2, x_3)$, where $x_2 \in \mathcal{X}_{m1}$ and $x_3 \in \mathcal{X}_{m2}$, resulting in four matrix equations with common components:
	\begin{eqnarray*}
		J_{c_1,x_3} &=& L_{X_{r_1}|c_1,\Xst} \Delta_{c_1,x_3,\Xst}H_{X_{r_4}|x_3,\Xst},\\
		J_{x_2,x_3} &=& L_{X_{r_1}|x_2,\Xst} \Delta_{x_2,x_3,\Xst}H_{X_{r_4}|x_3,\Xst},
	\end{eqnarray*}
	which share a common operator $H_{X_{r_4}|x_3,\Xst}$. 
	
	Similarly, we then have the following equations at two different values of $(X_{r_2}, X_{r_3})$, $(x_2,c_3)$ and $(c_1, c_3)$:
	\begin{eqnarray*}
		J_{x_2,c_3} &=& L_{X_{r_1}|x_2,\Xst} \Delta_{x_2,c_3,\Xst}H_{X_{r_4}|c_3,\Xst},\\
		J_{c_1,c_3} &=& L_{X_{r_1}|c_1,\Xst} \Delta_{c_1,c_3,\Xst}H_{X_{r_4}|c_3,\Xst},
	\end{eqnarray*}
	which share a common operator $H_{X_{r_4}|c_3,\Xst}$.
	
	We impose the following injective assumptions on all four operators:
	\begin{ass}(Injective)  \label{ass_injective2} there exists one division of the domain such that the operators $ L_{X_{r_1}|c_1,\Xst}$, $L_{X_{r_1}|x_2,\Xst}$, $H_{X_{r_4}|x_3,\Xst}$, and $H_{X_{r_4}|c_3,\Xst}$ are injective for $\mathcal{G}=\mathcal{L}^1$.
	\end{ass}
	
	With such an injective assumption being satisfied, we obtain the following main equation:
	\begin{eqnarray}
		J_{c_1,x_3} J^{-1}_{x_2,x_3} J_{x_2,c_3}J^{-1}_{c_1,c_3}=L_{X_{r_1}|c_1,\Xst}\Delta_{c_1,x_2,x_3,c_3}L^{-1}_{X_{r_1}|c_1,\Xst}, \label{eq_joint24}
	\end{eqnarray}
	where the left-hand side matrix  can be computed directly from the data, and the right-hand side matrix is the linear integral operator $L_{X_{r_1}|c_1,\Xst}$ with the conditional density, with diagonal operator $\Delta_{c_1,x_2,x_3,c_3}$ defined as
	$$\Delta_{c_1,x_2,x_3,c_3}= \Delta_{c_1,x_3,\Xst} \Delta^{-1}_{x_2,x_3,\Xst}\Delta_{x_2,c_3,\Xst}\Delta^{-1}_{c_1,c_3,\Xst}.$$
	
	Equation \eqref{eq_joint24} indicates that the operator $J_{c_1,x_3} J^{-1}_{x_2,x_3} J_{x_2,c_3}J^{-1}_{c_1,c_3}$ can be represented as an eigenvalue-eigenfunction decomposition for the unknown operators $L_{X_{r_1}|c_1,\Xst}$ and $\Delta_{c_1,x_2,x_3,c_3}$ being the eigenvalues and eigenfunctions, respectively. The eigenfunctions $L_{X_{r_1}|c_1,\Xst}$, indexed by the latent factor, provide the unobserved conditional densities of order statistic $X_{r_1:n}|X_{r_2:n}=c_1,\Xst$. 
	
	For unique decomposition, we further impose restrictions on the relationship between the observed measurement $X$ in segment $\mathcal{X}_m$ and the latent factor $\Xst$. Specifically,
	\begin{ass}(Distinct)  \label{ass_distinct2} There exists one division of the domain such that the set $\{(x_2,x_3):  \frac{f_{r_2,r_3:n}(c_1,x_3|\xst_1)f_{r_2,r_3:n}(x_2,x_3|\xst_1)}{f_{r_2,r_3:n}(x_2,c_3|\xst_1)f_{r_2,r_3:n}(c_1,c_3|\xst_1)} \neq  \frac{f_{r_2,r_3:n}(c_1,x_3|\xst_2)f_{r_2,r_3:n}(x_2,x_3|\xst_2)}{f_{r_2,r_3:n}(x_2,c_3|\xst_2)f_{r_2,r_3:n}(c_1,c_3|\xst_2)}, where ~ x_2 \in  \mathcal{X}_{m1} ~\& x_3 \in  \mathcal{X}_{m2}\}$, has positive probability for all $\xst_1, \xst_2 \in \XXst$ whenever $\xst_1 \neq \xst_2$.
	\end{ass}
	
	With both assumptions satisfied, we can identify operator $L_{X_{r_1}|c_1,\Xst}$ up to scales from Equation \eqref{eq_joint24} using eigenfunction decomposition. Additionally, we can identify the conditional density $f_{r_1|r_2:n}(x_{1}|c_1,\xst)$ using the fact that the identified operator is associated with this density. We further pin down the scales using the fact that $\int_{x_1 \in \mathcal{X}_l} f_{r_1|r_2:n}(x_{1}|c_1,\xst) dx_1 = 1$. Once the scales are pinned down, we identify the conditional density in segment ``low," i.e., $f_{r_1|r_2:n}(x_{1}|c_1,\xst)$ for all $x_1 \le c_1$. Note that the conditional distribution $f_{r_1|r_2:n}(x_{1}|c_1,\xst) $ is the same as the density of $r_1$th order statistics from a sample of size $(r_2 -1)$ based on the parent distribution that is truncated on the right at $c_1$, i.e., $\frac{f^X(x|\xst)}{F^X(c_1|\xst)}$. Therefore, we identify this truncated distribution $\frac{f^X(x|\xst)}{F^X(c_1|\xst)}$, indicating that we identify the parent density $f^X(x|\xst)$ in segment ``low" up to an unknown scale $F^X(c_1|\xst)$ for all $x_1 \le c_1$. 
	
	%
\paragraph{Step 2} Using the identified operator $L_{X_{r_1}|c_1,\Xst}$ in segment ``low," we first identify an operator defined to be associated with the density in both segments ``middle" and ``high" using the joint density of the first three OS: 
	\begin{eqnarray}
		f_{r_1,r_2,r_3:n}(x_{1}, x_{2},x_{3})&=& \int_{\xst} f_{r_1|r_2:n}(x_{1}|x_2,\xst) f_{r_3|r_2:n}(x_3|x_2, \xst) f_{r_2:n}(x_2,\xst) d \xst. \notag
	\end{eqnarray}
	
	We then exploit the equivalence of the linear integral operator when fixing $X_{r_2}= c_1$ to identify the conditional distribution. Particularly, we can derive the following operator equivalence for $X_{r_2}= c_1$ and any $X_{r_3}=y \ge c_1$. Specifically,
	\begin{eqnarray*}
		[J_{c_1}g](x_1) &\equiv & \int_{y \ge c_1} f_{r_1,r_2,r_3:n}(x_{1}, c_1,y) g(y)dy \notag \\
		&=& \int_{y \ge c_1}  \int_{\xst}  f_{r_1|r_2:n}(x_{1}|c_1,\xst) f_{r_3|r_2:n}(x_3|c_1, \xst) f_{r_2:n}(x_2,\xst) d \xst dy \notag  \\
		&=&  \int_{\xst} f_{r_1|r_2:n}(x_{1}|c_1,\xst)   f_{r_2:n}(x_2,\xst) \int_{y \ge c_1} f_{r_3|r_2:n}(y|c_1, \xst)g(y) dy d \xst \notag \\
		&=&  \int_{\xst} f_{r_1|r_2:n}(x_{1}|c_1,\xst) [f_{r_2:n}(x_2,\xst)  M_{X_{r_3}|c_1,\Xst} g](\xst) d\xst \notag \\
		&=& [L_{X_{r_1}|c_1,\Xst} \Delta_{c_1,\Xst} M_{X_{r_3}|c_1,\Xst}g](x_1),
	\end{eqnarray*}
	where $M_{X_{r_3}|c_1,\Xst} g](\xst)\equiv \int_{y \ge c_1} f_{r_3|r_2:n}(y|c_1, \xst)g(y)dy$ and  $[\Delta_{c_1,\Xst} g](\xst)\equiv f_{r_2:n}(x_2,\xst)g(\xst)$.
	We obtain the equivalence of operators in the following:
	\begin{eqnarray*}
		J_{c_1} = L_{X_{r_1}|c_1,\Xst} \Delta_{c_1,\Xst} M_{X_{r_3}|c_1,\Xst}.
	\end{eqnarray*}
	Therefore, we can identify the operator $M_{X_{r_3}|c_1,\Xst}$ up to scales, since the operator $L_{X_{r_1}|c_1,\Xst} $ is identified and operator $\Delta_{c_1,\Xst}$ is a diagonal. In addition, we identify the conditional density $f_{r_3|r_2:n}(x_3|c_1, \xst)$ up to scales. We can pin down the scales using the fact that $\int_{y \ge c_1} f_{r_3|r_2}(y|c_1,\xst)dx = 1$, so we identify fully $f_{r_3|r_2}(x_{3}|c_1,\xst)$ for all $x \ge c_1$. Note that the conditional distribution $f_{r_3|r_2}(x|c_1,\xst)$ is the same as the density of $(r_3-r_2)$th order statistics for a sample of size $(n-r_2)$ from a distribution that is truncated on the right at $c_1$, i.e., $\frac{f^X(x|\xst)}{1-F^X(c_1|\xst)}$. Therefore, we identify this truncated distribution $\frac{f^X(x|\xst)}{1-F^X(c_1|\xst)}$ and $f^X(x|\xst)$ up to an unknown scale $[1-F^X(c_1|\xst)]$ for all $x \ge c_1$. 
	
\paragraph{Step 3} To summarize, we identify the conditional distribution density $f^X(x|\xst)$ for $x \le c_1$ up to an unknown scale $F^X(c_1|\xst)$ for all $x_1 \le c_1$ and identify the conditional distribution $f^X(x|\xst)$ for $x\ge c_1$ up to an unknown scale $[1-F^X(c_1|\xst)]$. We then pin down the unknown $F^X(c_1|\xst)$ using the smoothness of the conditional density, i.e., 
	\begin{flalign*}
		\frac{f^X(x|\xst)}{F^X(c_1|\xst)} = \frac{f^X(x|\xst)}{1-F^X(c_1|\xst)},
	\end{flalign*}
	which admits a unique and explicit solution for $F^X(c_1|\xst)$. 
	
	Note that the identification argument described above generates a continuous conditional distribution without knowing the associated value of the latent true factor $\Xst$. The identification is up to ordering and location. Pinning down the exact value of such location calls for extra restrictions, which typically depend on the context of the latent factor. We use the same restriction as in the previous section to pin down the location. 
	
	We summarize the identification result in the following theorem. 
	
	\begin{theorem}
		If  Assumptions \ref{cond_id}, \ref{ass_bound}, \ref{ass_injective2},  \ref{ass_distinct2}, and \ref{ass_order} are satisfied, the conditional density distribution $f^{X}(x|\Xst)$ for $x \in \mathcal{X}$ and the marginal distribution for the latent variable $f^{\Xst}(\xst)$ for $\xst\in \XXst$ are identified using any four order statistics. 
	\end{theorem}
	
	
	\bibliographystyle{ecta}
	\bibliography{refOS}